\documentclass[usenatbib,useAMS,fleqn]{mnras}

\usepackage[T1]{fontenc}
\usepackage{ae,aecompl}

\usepackage{tikz, graphicx}
\usepackage{amsmath}	
\usepackage{amssymb}	
\usepackage{multicol}        
\usepackage{threeparttable}
\usepackage{xspace}
\usepackage{times}

\usepackage{newtxtext,newtxmath}

\usepackage{etoolbox}
\makeatletter
\patchcmd\@combinedblfloats{\box\@outputbox}{\unvbox\@outputbox}{}
\makeatother

\newcommand{\mic}{\,$\mu$m} 
\newcommand{\Ha}{H$\alpha$}
\newcommand{\HH}{H$_2$}
\newcommand{\HII}{H$\,${\sc ii}}     

\newcommand{\tstar}{$t_{\rm star}$}   
\newcommand{\tHa}{$t_{\rm H\alpha}$}   
\newcommand{\tgas}{$t_{\rm gas}$}   
\newcommand{\tCO}{$t_{\rm CO}$}   
\newcommand{\tover}{$t_{\rm fb}$}   
\newcommand{\tsn}{$t_{\rm SN}$} 
\newcommand{\tcomb}{$t_{\rm comb}$} 
\newcommand{\tmin}{$t_{\rm min}$} 
\newcommand{\esf}{$\epsilon_{\rm sf}$}   
\newcommand{\msun}{\mbox{M$_\odot$}}

\newcommand{\myr}{\mbox{${\rm Myr}$}}

\newcommand{\pc}{\mbox{${\rm pc}$}}

\newcommand{\cmc}{\mbox{${\rm cm}^{-3}$}}

\title[Early feedback drives cloud destruction]{Pre-supernova feedback mechanisms drive the destruction of molecular clouds in nearby star-forming disc galaxies}

\author[Chevance et al.]{M\'elanie Chevance,$^{1}$\thanks{Contact e-mail: chevance@uni-heidelberg.de} 
J.~M.~Diederik Kruijssen,$^{1}$
Mark R.~Krumholz,$^{2,3,4,5}$
\newauthor
Brent Groves,$^{6}$
Benjamin W.~Keller,$^{1}$
Annie Hughes,$^{7,8}$
Simon C.~O.~Glover,$^{9}$
\newauthor
Jonathan~D.~Henshaw,$^{5}$
Cinthya N.~Herrera,$^{10}$
Jenny J.~Kim,$^{1}$
Adam K.~Leroy,$^{11}$
\newauthor
J\'{e}r\^{o}me Pety,$^{10,12}$
Alessandro Razza,$^{13}$
Erik Rosolowsky,$^{14}$
Eva Schinnerer,$^{5}$
\newauthor
Andreas Schruba,$^{15}$
Ashley T.~Barnes,$^{16}$
Frank Bigiel,$^{16}$
Guillermo A. Blanc,$^{13,17}$
\newauthor
Eric Emsellem,$^{18,19}$
Christopher M. Faesi,$^{20}$
Kathryn Grasha,$^{2}$
Ralf S.~Klessen,$^{9,21}$
\newauthor
Kathryn Kreckel,$^{1}$
Daizhong Liu,$^{5}$
Steven N.~Longmore,$^{22}$
Sharon E.~Meidt,$^{23}$
\newauthor
Miguel Querejeta,$^{24}$
Toshiki Saito,$^{5}$
Jiayi Sun,$^{11}$
Antonio Usero$^{24}$ 
\\\\
Affiliations are listed at the end of the paper
}

\date{Accepted XXX. Received YYY; in original form 2020 October 23}

\pubyear{2020}

\begin{document}
\label{firstpage}
\pagerange{\pageref{firstpage}--\pageref{lastpage}}
\maketitle

\begin{abstract}
It is a major open question which physical processes stop the accretion of gas onto giant molecular clouds (GMCs) and limit the efficiency at which gas is converted into stars within  these GMCs. While feedback from supernova explosions has been the popular feedback mechanism included in simulations of galaxy formation and evolution, `early' feedback mechanisms such as stellar winds, photoionisation and radiation pressure are expected to play an important role in dispersing the gas after the onset of star formation. These feedback processes typically take place on small scales ($\sim 10-100$\,pc) and their effects have therefore been difficult to constrain in environments other than the Milky Way. We apply a novel statistical method to $\sim 1$\arcsec\ resolution maps of CO and \Ha\ emission across a sample of nine nearby disc galaxies, in order to measure the time over which GMCs are dispersed by feedback from young, high-mass stars, as a function of the galactic environment. We find that GMCs are typically dispersed within $\sim 3$\,Myr after the emergence of unembedded high-mass stars, showing no significant trend with galactocentric radius. Comparison with analytical predictions demonstrates that, independently of the environment, early feedback mechanisms (particularly photoionisation and stellar winds) play a crucial role in dispersing GMCs and limiting their star formation efficiency in nearby galaxies. Finally, we show that the efficiency at which the energy injected by these early feedback mechanisms couples with the parent GMC is relatively low (a few tens of per cent), such that the vast majority of momentum and energy emitted by the young stellar populations escapes the parent GMC.
\end{abstract}




\section{Introduction}

One of the main questions in studies of galactic star formation has been why the global depletion time (i.e.\ the time needed to consume the entire reservoir of molecular gas in a galaxy at the current star formation rate; SFR; e.g.~\citealt{Bigiel2008, Calzetti2012}) is two orders of magnitude larger than the time it would take the giant molecular clouds (GMCs) in which stars form, to collapse due to gravity in the absence of pressure support \citep[the dynamical time, see e.g.][]{Zuckerman1974, Murray2011, Evans2014, Vutisalchavakul2016, Leroy2017, Ochsendorf2017,Utomo2018, Schruba2019}.
To reconcile these two time-scales, two scenarios have been suggested \citep[see][for a recent review]{Chevance2020b}. In the first scenario, GMCs are not collapsing on a dynamical time, but are supported against gravitational collapse, e.g.\ by the presence of magnetic fields. As a result, the conversion of most of the molecular gas reservoir into stars takes place over many dynamical times \citep{McKee1989, Koda2009, VazquezSemadeni2011}. However, recent observational measurements of the magnetic field strength suggest that they are in general insufficient to support GMCs against collapse \citep{Crutcher2012, Crutcher2019}. The internal turbulence is also insufficient to provide persistent support to the clouds because, without replenishment, it dissipates on a cloud crossing time \citep{MacLow2004}.
In the second scenario, GMCs are transient objects which only survive for a dynamical time \citep{Elmegreen2000, Hartmann2001, Clark2005, Dobbs2011, Ward2016}, and only a small fraction of a GMC is actually converted into stars, while the vast majority of the gas is dispersed before the completion of the star formation process. In this case, the low efficiency of star formation on the cloud scale is responsible for the long galaxy-scale depletion time.

After a wide variety of early case studies \citep[e.g.][]{Kawamura2009, Miura2012, Meidt2015, Corbelli2017}, several recent observational studies have now shown systematically that GMCs and their progenitor H{\sc i} clouds live for about a dynamical time \citep[e.g.][]{Kruijssen2019, Chevance2020, HygatePhD, Ward2020_HI, Zabel2020} and are dispersed within a few Myr after the onset of high-mass star formation, as visible in \Ha\ emission \citep[e.g.][]{Whitmore2014, Hollyhead2015, Grasha2018, Grasha2019, Hannon2019, Chevance2020, Chevance2020b, Haydon2020b}. It is an open question which physical mechanisms drive GMC dispersal. While unbound clouds can potentially be dispersed by local dynamical processes or galactic shear \citep[e.g.][]{Dobbs2013,Dobbs2014,Jeffreson2018,Jeffreson2020}, stellar feedback additionally injects energy and momentum into the gas surrounding young stellar regions \citep[see e.g.][for a review]{Krumholz2019}, thereby driving the baryon cycle within the host galaxy. If strong enough, stellar feedback can potentially halt gas accretion and cloud collapse, and therefore limit the star formation efficiency. Determining which physical mechanism(s) disperse(s) the gas clouds in galaxies is therefore critical to understand what limits the efficiency of the conversion of gas into stars, and how this varies with environment (e.g.\ galaxy structure and morphology, gas and stellar surface densities, metallicity) across cosmic time.

It remains unclear which feedback mechanisms are dominant at each evolutionary phase of a young stellar region \citep[e.g.][]{Lopez2011, Lopez2014, McLeod2019, Barnes2020, Olivier2020}. Theoretical studies and numerical simulations have shown that feedback from supernovae alone is insufficient to disperse the dense gas, and that early feedback in the form of winds and radiation is crucial to limit the efficiency with which GMCs convert their gas into stars \citep[e.g.][]{Matzner2002, Krumholz2009, Agertz2013, Stinson2013, Dale2014, Hopkins2014, Dale2015, Gatto2015, Matzner2015, Geen2016, Gatto2017, Hu2017, Rahner2017, Rahner2019, Haid2018, Kim2018, Decataldo2020, Lucas2020}.
Stellar feedback mechanisms other than supernovae are only recently being added to simulations of galaxy formation and evolution, because these are beginning to resolve GMC scales \citep[e.g.][]{Grand2017, Hopkins2018}. However, the lack of observational constraints and the large number of free parameters make it difficult to accurately implement these forms of feedback in galaxy simulations \citep[e.g.,][]{Fujimoto2019}.

In a recent work, \citet{Chevance2020} used a statistical method to measure the durations of the successive phases of the GMC evolutionary lifecycle in a sample of nine nearby disc galaxies, mostly focusing on the environmental dependence of the molecular cloud lifetime. In this paper, we follow up on this work and use the same sample of nine galaxies to focus on the time over which molecular gas is dispersed after the first \Ha\ emission emitted by recently formed high-mass stars becomes detectable (which we refer to as the `feedback time-scale'). We compare these observational measurements with theoretical predictions for the characteristic time-scales associated with stellar feedback from supernovae, stellar winds, photoionisation and radiation pressure. This approach allows us to determine which processes play the dominant role in limiting the star formation efficiency of GMCs, as a function of the galactic environment. 

The structure of this paper is as follows. In Section~\ref{sec:measurement}, we describe the observational sample and the analysis method, present the global and resolved measured feedback time-scales in our galaxy sample, and validate the accuracy of these measurements. In Section~\ref{sec:predictions}, we compare the measured feedback time-scales with theoretical predictions for cloud dispersal by stellar feedback. Finally, we correlate these results with environmental properties and discuss their physical interpretation in Section~\ref{sec:discussion}. We conclude in Section~\ref{sec:conclusion}.

\section{Observational measurement of the feedback time-scale}
\label{sec:measurement}

In this section, we describe our observations and the analysis applied to the observed sample to measure the feedback time-scale. We then discuss the accuracy of our measurements.

\subsection{Observations}
\label{sec:observations}

We focus on a sample of nine star forming disc galaxies: NGC0628, NGC3351, NGC3627, NGC4254, NGC4303, NGC4321, NGC4535, NGC5068 and NGC5194. These galaxies were chosen to be relatively nearby (between 5\,Mpc and 18\,Mpc) to achieve sufficient spatial resolution (see Section~\ref{sec:requirements}), and be moderately inclined ($< 60^{\circ}$), which minimises the impact of dust attenuation along the line of sight or other projection effects. Details about the integrated properties of these galaxies, the data used and the associated reduction process can be found in \citet{Schinnerer2019} and \citet{Chevance2020}, and references therein. We summarise below the main characteristics of the observations used in this work.
To characterise the duration of the feedback time-scale as described in Section~\ref{sec:Heisenberg}, we use the emission from low-$J$ transitions of carbon monoxide [CO(1-0) and CO(2-1)], which is a common probe of the molecular gas mass \citep[e.g.][]{Bolatto2013, Sandstrom2013}, and \Ha\ emission, which is a common probe of the SFR \citep[e.g][]{Kennicutt2012}.

All galaxies presented here have been observed in CO(2-1) with the Atacama Large Millimeter/submillimeter Array \citep[ALMA; see][]{Leroy2020} as part of the PHANGS\footnote{Physics at High Angular Resolution in Nearby GalaxieS; \url{http://phangs.org}.}-ALMA survey, except for NGC5194, observed in CO(1-0) by the PAWS survey \citep{Pety2013, Schinnerer2013}. The angular resolution of these observations is $\sim 1$\arcsec\ ($\sim 35-160$\,\pc\ at the distances of these galaxies), which resolves the typical distance between independent star-forming regions (i.e.\ regions that reside on an evolutionary timeline at a phase that is independent of their neighbours, see e.g.\ \citealt{Kruijssen2014,Chevance2020,Tacchella2020}). The typical $3\sigma$ point source sensitivity of the CO observations corresponds to $\sim 10^5$\msun .

\Ha\ emission maps are also available at similar ($\sim 1$\arcsec) angular resolution for this galaxy sample, either from the Spitzer Infrared Nearby Galaxies Survey (SINGS; \citealt{Kennicutt2003}) or from new observations using the Wide-Field Interferometer (WFI) instrument on the 2.2-m MPG/ESO telescope at La  Silla  Observatory (Razza et al.\ in prep.). In the following, we use the term 'high-mass' star-forming regions to refer to those containing a sufficient number of high-mass stars ($\gtrsim 10$\,\msun ) to be detectable in \Ha . 

\subsection{Measurement of the feedback time-scale}
\label{sec:Heisenberg}

\begin{table*}
\begin{center}
\begin{tabular}{lllllllll}
\hline
                    NGC0628  &      Radial interval [kpc]  &                     entire  &                  0.77-2.58  &                  2.58-3.79  &                  3.79-5.00  &                  5.00-7.63  &                             &                             \\
                             &   Feedback time-scale [Myr]  &        $3.2^{+0.6}_{-0.4}$  &        $2.7^{+0.7}_{-0.9}$  &       $4.9^{+10.1}_{-1.0}$  &        $2.1^{+0.7}_{-0.7}$  &        $3.5^{+1.5}_{-0.8}$  &                             &                             \\[2.5ex]
                    NGC3351  &      Radial interval [kpc]  &                     entire  &                  2.34-3.50  &                  3.50-4.67  &                  4.67-6.14  &                             &                             &                             \\
                             &   Feedback time-scale [Myr]  &        $2.5^{+0.8}_{-0.6}$  &        $\leq3.8$  &        $2.6^{+0.9}_{-0.8}$  &        $1.5^{+2.0}_{-0.8}$  &                             &                             &                             \\[2.5ex]
                    NGC3627  &      Radial interval [kpc]  &                     entire  &                  0.69-2.66$^{\star}$  &                  2.66-3.68$^{\star}$  &                  3.68-4.70  &                  4.70-5.73  &                  5.73-8.78  &                             \\
                             &   Feedback time-scale [Myr]  &        $2.8^{+0.8}_{-0.7}$  &        $2.1^{+1.3}_{-0.8}$  &        $\leq 1.4$  &       $\leq 19.3$  &        $\leq 8.1$  &        $4.4^{+1.9}_{-1.3}$  &                             \\[2.5ex]
                    NGC4254  &      Radial interval [kpc]  &                     entire  &                  0.53-2.60  &                  2.60-4.25  &                  4.25-6.06  &                  6.06-7.86  &                  7.86-9.67  &                 9.67-13.77  \\
                             &   Feedback time-scale [Myr]  &        $4.8^{+1.1}_{-1.0}$  &        $\leq 9.3$  &        $3.4^{+2.4}_{-1.2}$  &        $3.7^{+1.5}_{-1.0}$  &      $\leq 22.2$  &        $3.6^{+1.3}_{-1.3}$  &        $4.2^{+1.4}_{-1.5}$  \\[2.5ex]
                    NGC4303  &      Radial interval [kpc]  &                     entire  &                  1.16-3.10$^{\star}$  &                  3.10-4.39$^{\star}$  &                  4.39-5.68  &                  5.68-6.97  &                  6.97-9.50  &                             \\
                             &   Feedback time-scale [Myr]  &        $4.0^{+1.7}_{-1.0}$  &        $2.1^{+1.3}_{-0.9}$  &        $4.5^{+4.9}_{-1.6}$  &        $4.9^{+5.1}_{-1.4}$  &        $4.2^{+5.7}_{-1.5}$  &        $1.5^{+1.0}_{-0.7}$  &                             \\[2.5ex]
                    NGC4321  &      Radial interval [kpc]  &                     entire  &                  0.95-4.18$^{\star}$  &                  4.18-5.71$^{\star}$  &                  5.71-7.24  &                  7.24-8.77  &                 8.77-10.31  &                10.31-13.54  \\
                             &   Feedback time-scale [Myr]  &        $3.3^{+0.7}_{-0.6}$  &        $3.0^{+2.6}_{-1.1}$  &        $2.8^{+1.1}_{-0.8}$  &        $3.2^{+1.5}_{-0.9}$  &        $4.8^{+0.9}_{-1.6}$  &        $\leq 2.1$  &        $5.5^{+5.3}_{-2.3}$  \\[2.5ex]
                    NGC4535  &      Radial interval [kpc]  &                     entire  &                  3.02-5.09$^{\star}$  &                  5.09-7.06  &                 7.06-10.98  &                             &                             &                             \\
                             &   Feedback time-scale [Myr]  &        $3.9^{+1.2}_{-0.9}$  &      $\leq 27.0$  &        $3.6^{+1.0}_{-0.6}$  &        $2.6^{+1.3}_{-1.0}$  &                             &                             &                             \\[2.5ex]
                    NGC5068  &      Radial interval [kpc]  &                     entire  &                  0.00-1.62  &                  1.62-2.70  &                  2.70-5.18  &                             &                             &                             \\
                             &   Feedback time-scale [Myr]  &        $1.0^{+0.4}_{-0.3}$  &        $1.7^{+0.5}_{-0.3}$  &        $1.3^{+1.4}_{-0.5}$  &        $\leq 0.4$  &                             &                             &                             \\[2.5ex]
                    NGC5194  &      Radial interval [kpc]  &                     entire  &                  0.51-1.77  &                  1.77-2.93  &                  2.93-4.09  &                  4.09-5.35  &                             &                             \\
                             &   Feedback time-scale [Myr]  &        $4.8^{+2.1}_{-1.1}$  &        $3.3^{+3.6}_{-1.2}$  &      $\leq 20.5$  &        $2.2^{+2.2}_{-0.8}$  &        $6.6^{+4.9}_{-2.0}$  &                             &                             \\[2.5ex]
\hline
\end{tabular}
\caption{Measured feedback time-scale, \tover , for each galaxy in its entirety, as well as in each individual radial bin. Bins containing a bar or the end of a bar are indicated with a $^{\star}$. These regions are susceptible to bursty episodes of star formation, which may affect our measurements (see Section~\ref{sec:morphology}). For bins affected by blending, only an upper limit on \tover\ is given (see Section~\ref{sec:requirements}).}
\label{tab:fb}
\end{center}
\end{table*}

Resolved observations of nearby galaxies have shown that the galaxy-scale star formation relation linking the gas surface density to the SFR surface density \citep[e.g.][]{Kennicutt1998} breaks down below a few 100\,pc \citep[e.g.][]{Bigiel2008, Onodera2010, Schruba2010, Leroy2013, Kreckel2018, Kruijssen2019, Schinnerer2019, Chevance2020}. The observed scatter on these sub-kpc scales has been interpreted as a sign of the evolutionary cycling of independent star-forming regions, from cloud assembly, to cloud collapse, star formation and eventually cloud destruction by stellar feedback \citep[e.g.][]{Schruba2010, Feldmann2011, KL14}. On the small scales of individual GMCs or \HII\ regions, a specific region is observed at a given time in this cycle. If observed at the stage of a non-star forming GMC, the local gas-to-SFR ratio will be high relative to the average value for that galaxy. By contrast, if observed at a later stage, during the young, unembedded star-cluster phase, after gas dispersal, the local gas-to-SFR ratio will be low relative to the average value. The tight star formation relation observed on large scales therefore results from averaging over many independent regions, which each individually sample this timeline. The link between the observed scatter of the gas-to-SFR ratio on the cloud scale ($\sim 100$\,pc) and the duration of the different phases of this cycle has been recently formalised by \cite{KL14} and \cite{Kruijssen2018}.

The total duration of the evolutionary cycle between molecular clouds, star formation and feedback can be described as:
\begin{equation}
\tau =  t_{\rm gas} + t_{\rm star} - t_{\rm fb},
\end{equation}
where \tgas\ is the lifetime of molecular clouds, \tstar\ is the duration of the young stellar phase, and \tover\ the duration of the overlap between these two phases (i.e.\ during which gas and young stars coexist). In the following, we will refer to the duration of the overlap phase as the `feedback time-scale'. This interpretation is discussed in Section~\ref{sec:interpretation}. The `\textit{gas}' and `\textit{star}' phases can be observationally probed by specific tracers. Here, we use CO emission\footnote{We note that the choice of the CO-to-\HH\ conversion factor does not affect the time-scales determined here, unless there are considerable variations within the galaxy on the scale of independent regions \citep[a few 100\,pc; see also discussion in][]{Kruijssen2018, Chevance2020}. In addition, in the galaxies of our sample, ranging from solar to half-solar metallicity, we do not expect high-mass star formation to take place in completely CO-dark clouds, which would require a different calibration of the timeline \citep{Haydon2020b}.} and \Ha\ emission to trace molecular clouds and young high mass (unembedded) stellar regions, respectively (see Section~\ref{sec:observations}) and measure \tCO\ = \tgas\ and \tHa\ = \tstar\ using the statistical method presented by \cite{Kruijssen2018} and the timeline calibration ($t_{\rm H\alpha} - t_{\rm fb} = 4.32$\,Myr at solar metallicity) derived by \cite{Haydon2020a}, using stellar population synthesis to determine the duration of the isolated \Ha\ emission phase. In practice, we measure the CO-to-\Ha\ flux ratio in apertures centred on CO (resp. \Ha) emission peaks for a series of aperture sizes (ranging from the resolution of the maps to $\sim$\,kpc sizes), and fit the relative excess (resp. deficit) of CO-to-\Ha\ flux ratio compared to the galactic averaged value, as a function of the aperture size. The fitted model depends on three free parameters: \tCO , \tover, and $\lambda$, where the latter is the characteristic distance between independent regions. These parameters, their associated uncertainties, as well as secondary quantities such as the integrated star formation efficiency per star formation event (defined as the ratio of the mass of the formed stars during a cloud lifetime and the gas mass) and the feedback outflow velocity are derived self-consistently. We refer the reader to section 3 of \cite{Kruijssen2018} for more details about the method and to \cite{Chevance2020} for a description of the input parameters and the general results for the cloud lifetimes, characteristic distance between regions, star formation efficiency and feedback outflow velocity in the sample of galaxies analysed here. In summary, for the sample of galaxies considered here, we measure a range of cloud lifetimes $t_{\rm CO} \sim 10-30$\,Myr, characteristic distances between regions $\lambda \sim 100-250$\,pc, and star formation efficiencies $\epsilon_{\rm sf} \sim 4-10$\,per cent.

The constrained feedback time-scale, which quantifies here the duration for the disruption of the CO clouds by stellar feedback from high-mass stars (see discussion in Section~\ref{sec:interpretation}), is presented in Table~\ref{tab:fb} for each of the nine galaxies of our sample, as well as for individual radial bins (of width $\sim 1-4$\,kpc) within these galaxies. We adopt the same radial bins as in \citet{Chevance2020}, designed to have a minimum width of 1\,kpc and a minimum number of 50 emission peaks in each tracer. On average, for full galaxies, the feedback time-scale is well constrained and relatively short, between 1.0\,Myr and 4.8\,Myr. Similarly short time-scales of a few Myr for GMC dispersal have been observed by previous studies \citep{Kawamura2009, Whitmore2014,Hollyhead2015,Corbelli2017,Grasha2019,Hannon2019, Kruijssen2019, HygatePhD}. With the exception of NGC5068 (which is a low-mass galaxy, with a low-pressure ISM, and relatively low signal-to-noise CO detections), there is little variation between galaxies, with \tover\ ranging between 2.5\,Myr and 4.8\,Myr. We note that larger variations (up to $\sim $7\,Myr) are observed between individual bins. These bin-to-bin variations are not always physical and may result from methodological biases, as discussed in Section~\ref{sec:predictions}.

For galaxy-averaged values, the feedback time-scale is shorter than the typical minimum delay time of the first supernova explosion \citep[4\,Myr; e.g.][]{Leitherer14} for five galaxies in our sample (NGC628, NGC3351, NGC3627, NGC4321 and NGC5068). For the other four galaxies, the feedback time-scale is consistent with this minimum delay time.
This short duration of the overlap phase between CO and \Ha\ suggests that early feedback mechanisms, such as photoionisation, stellar winds or radiation pressure must play an important role in disrupting the parent molecular cloud, only a few Myr after the emergence of unembedded high-mass stars.
We explore this hypothesis by comparing the measured \tover\ with theoretical predictions for different feedback mechanisms in Section~\ref{sec:predictions}.

\subsection{Accuracy of the results}
\label{sec:requirements}

\begin{figure}
	\includegraphics[width=\columnwidth]{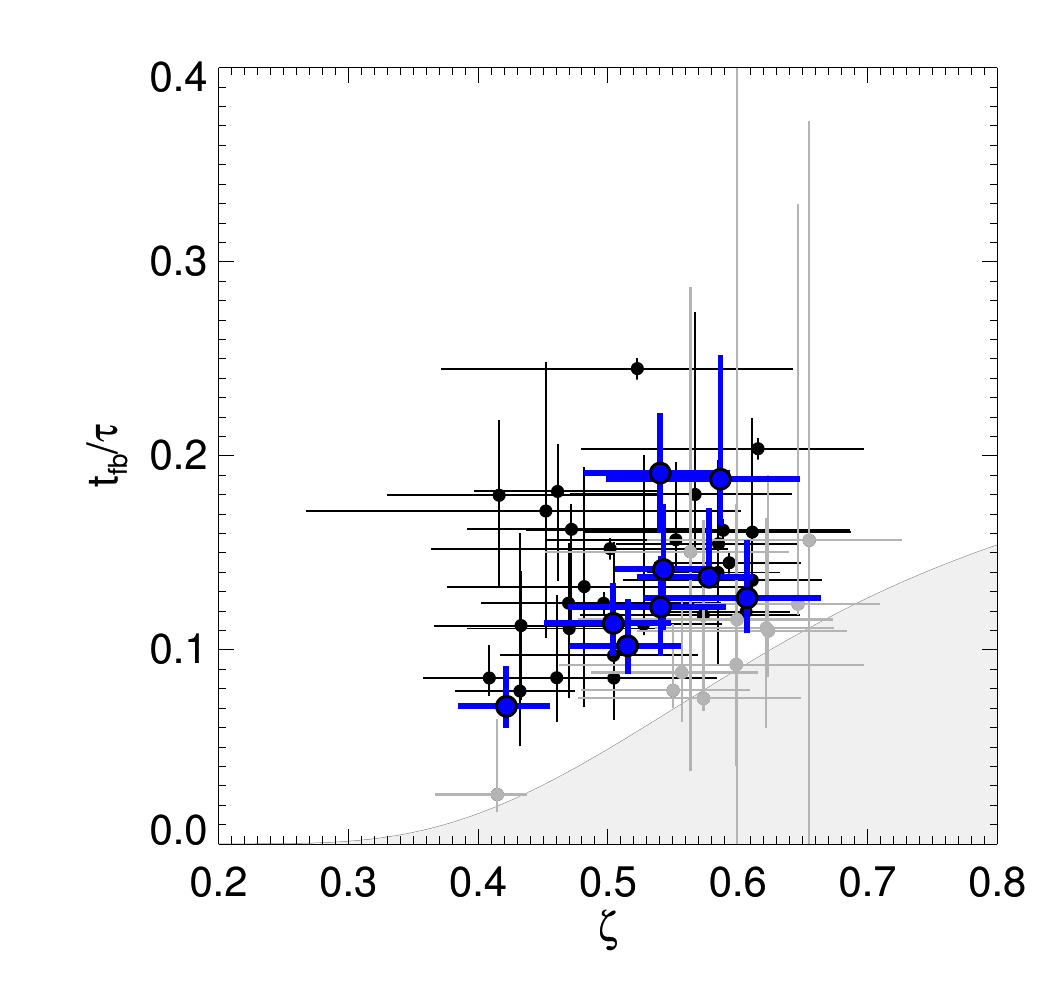}%
    \caption{Measured fraction of the evolutionary timeline spent in the feedback phase (\tover /$\tau$) as a function of the region filling factor $\zeta$, calculated as the maximum between the gas and the young stellar filling factors. Each thin data point represents the measurement for one radial bin. Thick blue data points represent galaxy-wide measurements. The grey-shaded area indicates the part of parameter space where measurements are affected by blending (at high filling factors and low feedback time-scales) as calculated in Appendix B of \citet{Kruijssen2018}}. The grey-shaded data points are $1\sigma$-consistent with being under the predicted curve and are therefore susceptible to blending. Overall, 10 (radial bin) data points out of 48 measurements fall in this region and therefore only provide upper limits on \tover .
    \label{fig:blending}
\end{figure}

To validate the accuracy of our measurements, we verify here that we fulfil the requirements listed in section 4.4 of \citet{Kruijssen2018}.
In \cite{Chevance2020}, we discussed the applicability of this method to CO and \Ha\ emission maps by verifying the requirements that the durations of the gas and stellar phases do not differ by more than an order of magnitude, that the typical distance between clouds is resolved, that the number of detected peaks per map and per radial bin is above 35 (satisfied by definition due to our choice of bins), and that any reservoir of diffuse emission is successfully filtered.
This means that our observations satisfy requirements (i) to (vi) listed in \citet{Kruijssen2018}, which guarantee that the measured values of \tCO , \tover\ and $\lambda$ are not biased.

For feedback-related quantities (i.e.\ \tover\ and quantities depending on it), additional requirements apply and are listed below. 
\begin{enumerate}
\item[(vii)] We ensure that our choices of parameters used in the emission peak identification process (listed in Table 3 of \citealt{Chevance2020}) enables the correct identification of adjacent peaks. Small variations of theses parameters do not affect our measurements.
\item[(viii)] The temporal overlap measured by \tover\ can be artificially increased by spatial overlap due to blending between individual regions. \citet{Kruijssen2018} specify constraints on the ratio between the feedback time-scale and the total duration of the evolutionary cycle, \tover /$\tau$, and on the region filling factor $\zeta = r/\lambda$ (where $r$ is the average radius of emission peaks) for the measurements not to be affected by blending. To verify that our data satisfy this condition, we place our measurements in the plane \tover /$\tau$ versus $\zeta$ in Figure~\ref{fig:blending}. The grey-shaded area indicates the part of the parameter space where measurements are affected by blending. Across the 48 measurements presented here, 10 radial bins overlap with this region within their $1\sigma$ uncertainties (grey data points in Figure~\ref{fig:blending}), so that we measure only an upper limit on the feedback time-scale in these radial bins. These data are indicated as such in Table~\ref{tab:fb}. We note that blending only affects the measurement of feedback-related quantities and that \tCO\ and $\lambda$ are still determined with good accuracy in these cases.
\item[(ix)] In all radial bins (except for the outermost bin of NGC5068, which is affected by blending) and full galaxies, we verify $0.05 < t_{\rm fb}/\tau < 0.95 $, which ensures good precision on the feedback time-scale (though it remains possibly limited according to point (viii) above).
\item[(x)] The star formation history should not exhibit significant variations over the total duration of the evolutionary timeline. While the detailed star formation histories of the galaxies in our sample are not known, we assume that there are no significant variations of their SFR (by more than 0.2 dex) in the disc during the last $\tau$ (the duration of an entire cycle of star formation, i.e.\ $\sim$ 20 to 50\,Myr), when averaged over time intervals of a duration \tover . We note two possible cases where this assumption could be incorrect. The radial bins including a bar, or the end of a bar, or those located at the co-rotation radius are likely to have a bursty star formation history, where star formation events are synchronised in time and space due to the presence of these large-scale morphological features (galactic centres, which can also have a bursty star formation are already excluded from our analysis, see \citealt{Chevance2020}). This type of synchronisation violates the key assumption in our analysis method that the ages of regions homogeneously sample the underlying timeline. We therefore indicate the presence of a bar with a $^{\star}$ in Table~\ref{tab:fb}. Where appropriate, we clearly indicate radial bins that are potentially affected by a bar or co-rotation in the subsequent figures. Another possible exception to a constant star formation history is the galaxy NGC3627, which has a pronounced bar \citep{Beuther2017}, shows complex molecular gas motions (Be\v{s}li\'c et al.\ in prep.) and is in interaction with the neighbouring galaxy NGC3628, and for which the star formation history may have varied recently \citep[e.g.][]{Rots1978, Haynes1979}. Despite also being in an interacting system, M51 seems to have a relatively constant star formation history within the past 100\,Myr \citep{Eufrasio2017}. 
\item[(xi)] Visual inspection of the maps does not reveal any remaining, abundant region blending in the areas were the analysis is performed (after excluding galactic centres; see \citealt{Chevance2020}).
\end{enumerate}
In conclusion, with the exception of the radial bins affected by blending or the presence of a bar (15 out of 48 measurements), the analysis performed here fulfils the requirements from \citet{Kruijssen2018} for feedback-related quantities. This validates the accuracy of our measurements.

\section{Variation of the feedback time-scale as a function of galactic environment}
\label{sec:predictions} 

In this section, we discuss how the measured feedback time-scales vary with galactic properties and morphology, and compare these measurements with theoretical predictions for stellar feedback mechanisms capable of destroying GMCs: supernova explosions, photoionisation, stellar winds and radiation pressure. Other feedback mechanisms (e.g.\ protostellar outflows) act on sub-cloud scales and lack the power to disrupt entire GMCs \citep{Bally2016, Klessen2016, Krumholz2019} and their effects and characteristic time-scales are therefore not investigated here.

\subsection{Environmental dependence}
\label{sec:environment}

\begin{figure*}
	\includegraphics[width=\linewidth]{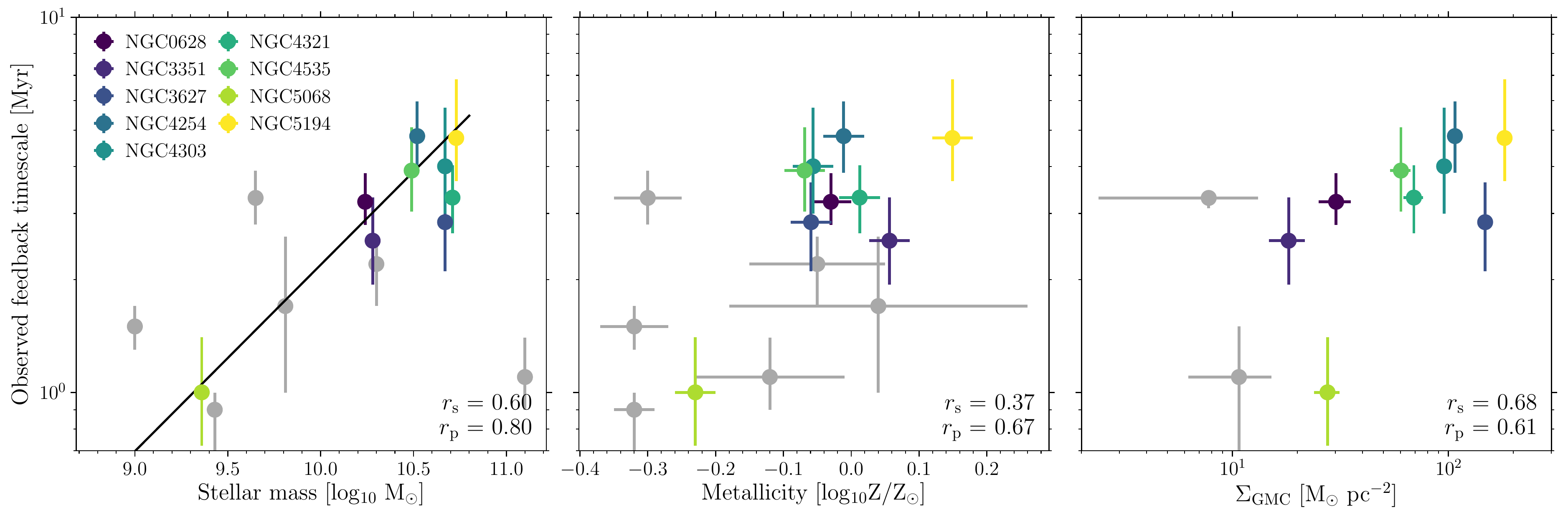}%
    \caption{Measured feedback time-scale as a function of the galaxy stellar mass (left), the average gas phase metallicity relative to solar (middle) and the median GMC surface density in the disc of the galaxy measured at 120\,pc resolution by \citet[right]{Sun2018}. The solar metallicity is defined as $12 + \log({\rm O}/{\rm H}) = 8.69$ by \citet{Asplund2009}. The observed feedback time-scale increases with stellar mass, metallicity and GMC surface density across the galaxy sample considered here, with Spearman and Pearson correlation coefficients ($r_{\rm s}$ and $r_{\rm p}$, respectively) as indicated in the bottom right corner of each panel. The grey data points show results for other nearby galaxies from the literature (see the text and Table~\ref{tab:literature}), and are not included in the calculation of the correlation coefficients.}
    \label{fig:tfb_environment}
\end{figure*}

We explore potential correlations between the feedback time-scale and environmental properties. In Figure~\ref{fig:tfb_environment}, we show the galaxy-wide feedback time-scale (i.e.\ measured across the full field of view) as a function of the galaxy stellar mass $M_{\star}$, the galaxy gas-phase metallicity and the median GMC surface density in the galaxy disc measured at 120~pc resolution $\Sigma_{\rm GMC}$ \citep{Sun2018}. \citet{Sun2018} assume a constant value of the CO-to-\HH\ conversion factor of $\alpha_{\rm CO(1-0)}=4.35$\,M$_{\odot}$\,pc$^{-2}$\,(K\,km\,s$^{-1}$)$^{-1}$ at solar metallicity, following \citet{Bolatto2013}. For consistency, we scale the value of $\Sigma_{\rm GMC}$ from \citet{Sun2018} with the average gas-phase metallicity measured in each galaxy. Similarly to \cite{Chevance2020}, we calculate the mean gas mass-weighted metallicity using the metallicity gradients determined by \citep{Pilyugin2014}, for all galaxies in our sample except NGC3627. For NGC3627, we use the metallicity gradient determined by \citep{Kreckel2019}, scaling the absolute value to compensate the different calibration method used in these two studies, based on the average metallicity of galaxies present in both samples.

The correlations shown in Figure~\ref{fig:tfb_environment} yield Spearman and Pearson correlation coefficients of \{0.60, 0.80\}, \{0.37, 0.67\} and \{0.68, 0.61\}, respectively. This means that the feedback time-scales are at least moderately correlated with the galaxy stellar mass, galaxy gas phase metallicity, and GMC surface density. The strongest correlation arises with stellar mass ($r_{\rm p}=0.80$). The correlation with metallicity is tentative. This might indicate that metallicity is not a fundamental parameter driving the evolution of the feedback time-scale, and could simply reflect the galaxy mass-metallicity relation \citep[e.g.][]{Tremonti2004}. Similarly, we find that the GMC surface density tightly correlates with stellar mass in our galaxy sample, suggesting that the GMC surface density is also not a fundamental parameter setting the feedback time-scale.

In the first two panels of Figure~\ref{fig:tfb_environment}, we additionally compare the results for our galaxy sample with other measurements of the feedback time-scale from the literature, for which the same analysis method as described in Section~\ref{sec:Heisenberg} was used. Table~\ref{tab:literature} summarises these literature results and the corresponding references for this compilation. The correlation between the measured feedback time-scale and the stellar mass persists when adding these additional data points, while the tentative correlation between the feedback time-scale and galaxy metallicity is more uncertain. In the third panel, we only add M31 and M33 from the literature compilation to ensure that the GMC surface density measurements are homogeneous, because the other literature galaxies are not included by \citet{Sun2018}.

\begin{table*}
\begin{center}
\begin{tabular}{lccccl}
\hline
Galaxy & \tover & Stellar mass & Metallicity & $\Sigma_{\rm GMC}$ & References \\
& Myr & $\log_{10}$\,M$_{\odot}$ & $\log_{10}$\,Z/Z$_{\odot}$ & $\log_{10}$\,M$_{\odot}$\,pc$^{-2}$ & \\
\hline
IC342 & $2.2^{+0.4}_{-0.5}$ & 10.3 & $-0.05\pm 0.10$ & -- & \cite{Jarrett2013, Kim2020} \\
LMC & $0.9^{+0.1}_{-0.2}$ & 9.4 & $-0.32\pm 0.03$ & -- & \cite{Besla2015, Kim2020} \\
M31 & $1.1^{+0.3}_{-0.2}$ & 11.1 & $-0.12\pm 0.11$ & $1.03 \pm 0.65$ & \cite{Tamm2012, Sun2018, Kim2020} \\
M33 & $3.3^{+0.6}_{-0.5}$ & 9.7 & $-0.30\pm 0.05$ & $0.89 \pm 0.73$ & \cite{Corbelli2003, Sun2018, Kim2020} \\
NGC300 & $1.5^{+0.2}_{-0.2}$ & 9.0 & $-0.32\pm0.05$ & -- & \cite{Westmeier2011, Kruijssen2019} \\
NGC1436 & $1.7^{+0.9}_{-0.7}$ & 9.8 & $0.04\pm0.22$ & -- & \cite{Zabel2020} \\
\hline
\end{tabular}
\caption{Compilation of feedback time-scale measurements from the literature, for which the same analysis method as described in Section~\ref{sec:Heisenberg} was used. The galaxy stellar masses, average metallicities and median GMC surface densities when available are indicated, with the literature references.}
\label{tab:literature}
\end{center}
\end{table*}

After the addition of the literature data, the tight correlation between the measured feedback time-scale and the stellar mass in the left-hand panel of Figure~\ref{fig:tfb_environment} remains. There is only one clear outlier, M31, with a mass just over $10^{11}$\,\msun. It is unclear if this galaxy is a true outlier (possibly linked to the fact that the analysed field of view covers only the outer parts of the galaxy, between 6 and 13\,kpc), or if the relation between the feedback time-scale and the stellar mass breaks above $M_{\star}\sim 10^{10.5}$\,\msun , similarly to the cosmic baryon efficiency \citep{Moster2013, Behroozi2013, Moster2018, Behroozi2019}.\footnote{It is an intriguing hypothesis that a possible peak of the feedback time-scale as a function of galaxy mass might be causally related to the peak in the baryon conversion efficiency. This could happen if the longest feedback time-scales result in the highest star formation efficiencies. While we cannot test this idea definitively given that we presently only have a measurement for M31 at stellar masses $M_{\star}>10^{10.75}$\,\msun, it will be straightforward to investigate this further once the full PHANGS-ALMA sample has been analysed with our statistical method.} We therefore exclude M31 and fit a power law function to the rest of the sample (including the literature data), for stellar masses between 9.0\msun\ and 10.8\msun , using an orthogonal distance regression \citep{Boggs1990}. This results in:
\begin{equation}
    \frac{t_{\rm fb}}{\rm Myr} = (2.19 \pm 0.25) \left(\frac{M_{\star}}{10^{10}M_{\odot}}\right)^{0.50 \pm 0.10}
\label{eq:fit}
\end{equation}
and the best fit is shown in the left-hand panel of Figure~\ref{fig:tfb_environment}. We measure an intrinsic scatter around this power law, defined as the residual after subtracting the scatter expected from the mean feedback time-scale error bars ($\sigma_{\rm err}$) from the standard deviation around the fit ($\sigma_{\rm meas}$), of $\sigma_{\rm intr} = (\sigma_{\rm meas}^2 - \sigma_{\rm err}^2)^{1/2} = 0.16$~dex, or 45~per~cent.
 
We expect the expansion of our analysis to the entire PHANGS-ALMA sample of $\sim80$ nearby galaxies (Kim et al.\ in prep.) to greatly improve the constraints on this potential relation, particularly in the high-mass regime.

\subsection{Influence of galaxy morphology}
\label{sec:morphology}

\begin{figure*}
	\includegraphics[width=\linewidth]{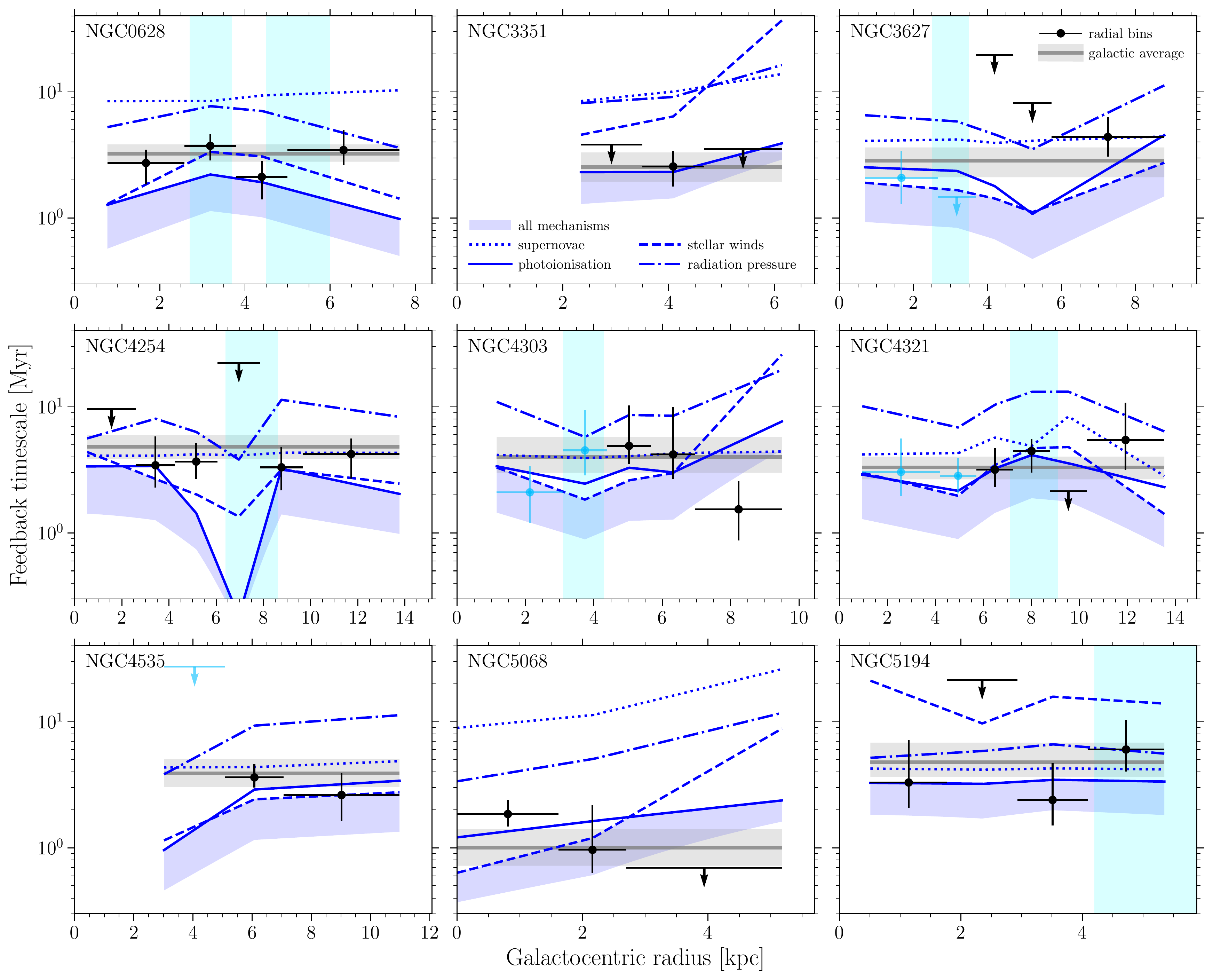}%
    \caption{Radial profiles of the observed feedback time-scale for each galaxy. The horizontal solid grey line represents the galactic average value, with the 1$\sigma$ uncertainties indicated by the grey-shaded area. For each bin of galactocentric radius, the horizontal error bar represents the range of radii over which the feedback time-scale is measured and the vertical error bar represents the 1$\sigma$ uncertainty on the measured value. For bins affected by blending (see the text and Figure~\ref{fig:blending}), only an upper limit on the feedback time-scale is shown. The cyan data points correspond to bins in the barred galaxies NGC3627, NGC4303, NGC4321 and NGC4535 that either include the bar or the end of the bar. The locations of the co-rotation radii \citep[][and references therein]{Chevance2020} are indicated by vertical cyan shaded areas, where the width indicates the uncertainty. The dark blue lines indicate the theoretical predictions for the supernova delay time (dotted line), as well as the time-scales for cloud dispersal by photoionisation (solid line), stellar winds (dashed line) and radiation pressure (dashed-dotted line), calculated as detailed in Section~\ref{sec:analytical}. In each panel, the blue-shaded area under the minimum of the blue lines indicates the range between the minimum of all predicted time-scales and the linear combination of all dispersal time-scales (\tcomb , excluding the supernova delay time, see the text).}
    \label{fig:profiles}
\end{figure*}

On the scale of $\sim$\,kpc radial bins, the measured feedback time-scale shows variations between and within galaxies (see Table~\ref{tab:fb} and Figure~\ref{fig:profiles}, where we show the measured feedback time-scales as a function of galactocentric radius). In radial bins including a bar, the end of a bar, or located at the co-rotation radius (indicated in Figure~\ref{fig:profiles} by cyan data points and vertical cyan shaded areas, respectively), the likely accumulation of gas and the associated bursty star formation can lead to locally enhanced or suppressed star formation \citep[e.g.][]{Beuther2017, Herrera2020}. This violates the requirement of an approximately constant SFR for the applicability of this method (see point (x) of Section~\ref{sec:requirements}), and likely explains some of the outliers observed in Figure~\ref{fig:profiles}. We note that in most cases these same bins are also affected by blending (e.g. NGC3627, NGC4535), which means that only an upper limit on the feedback time-scale can be estimated (see point (viii) of Section~\ref{sec:requirements} and Figure~\ref{fig:blending}). 

With the exception of bins affected by blending or morphological features, where methodological problems might affect the measurements, we note that the feedback time-scale shows no significant dependence on the galactocentric radius within galaxies. Although we note that the number of bins in which the feedback time-scale can be reliably measured is relatively limited in some of the galaxies of our sample, we will thus approximate the feedback time-scale as roughly constant in galaxies and will only consider the galaxy-wide measurements in the discussion section below (Section~\ref{sec:discussion}).

\subsection{Comparison with analytical predictions}
\label{sec:analytical}

We now compare the measured feedback time-scales to those predicted for a variety of feedback mechanisms. That way, we constrain which mechanisms drive GMC dispersal in our galaxy sample.

    \subsubsection{Supernova delay time}
\begin{figure}
	\includegraphics[width=\columnwidth]{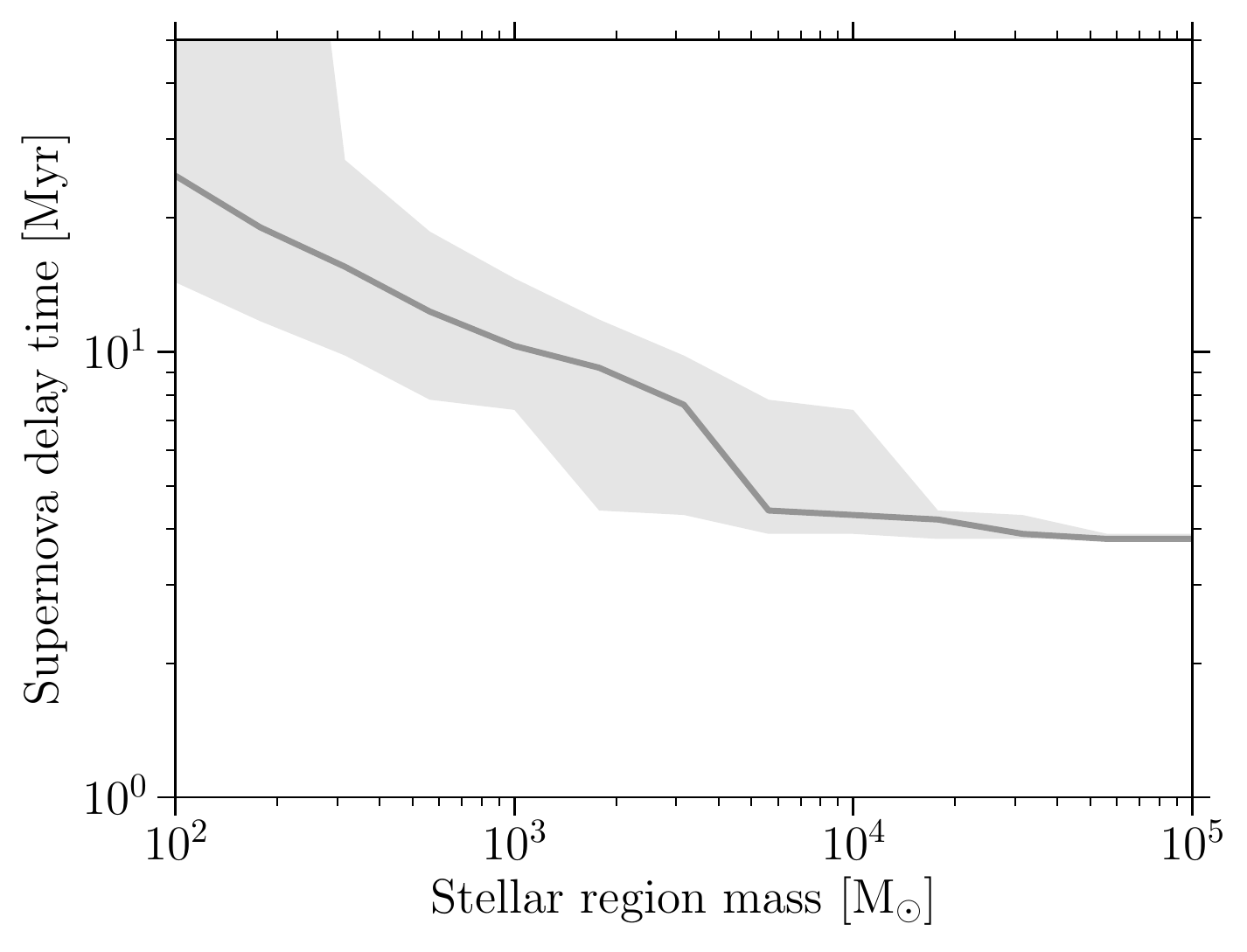}%
    \caption{Supernova delay time as a function of the stellar region mass. The grey shaded area shows the 16$^{\rm th}$-to-84$^{\rm th}$ percentile range of this value over $10^4$ Monte-Carlo realisations obtained with {\sc Slug}.}
    \label{fig:SN}
\end{figure}

Stellar evolution models predict that the delay time before the first supernova explosion is at least $\sim4$\,Myr in stellar regions where the initial stellar mass function (IMF) is well-sampled \citep{Leitherer14}. In the case of moderately low-mass stellar regions ($\lesssim 10^4$\,\msun), where the IMF is not fully sampled, this delay time is expected to be longer on average. Figure~\ref{fig:SN} shows the average supernova delay time as a function of the stellar region mass, predicted by the stochastic stellar population synthesis code SLUG \citep{Krumholz2015}, using a \cite{Chabrier2005} IMF and the non-rotating MESA Isochrones and Stellar Tracks (MIST; \citealt{Choi2016, Dotter2016}) at solar metallicity. We note that although the supernova delay time depends weakly on metallicity, the predicted delay times do not differ significantly from that at solar metallicity for the range of metallicities spanned by our galaxy sample. The mass range at which stars explode as supernovae (rather than collapsing directly to black holes) is taken from \cite{Sukhbold2016}.

For each observed radial bin in our galaxy sample, we determine the average stellar mass within an \HII\ region $M_\text{H{\sc ii}}$, expected from the average gas mass enclosed in a region of diameter $\lambda$ (which is the characteristic distance between independent regions; see Section~\ref{sec:Heisenberg}) multiplied by the (integrated) star formation efficiency \esf\ measured using our formalism \citep{Chevance2020}:
\begin{equation}
\label{eq:MHII}
M_\text{H{\sc ii}} = \pi \left(\frac{\lambda}{2}\right)^2 \epsilon_{\rm sf} \Sigma_{\rm gas}
\end{equation}
We then use the results of our SLUG calculation (Figure~\ref{fig:SN}) to estimate the expected supernova delay time for the average star-forming region in each radial bin. The predicted supernova delay time \tsn\ ranges between $\sim 4$ and 20\,Myr across our galaxy sample and is shown as a function of galactocentric radius in Figure~\ref{fig:profiles}.

     \subsubsection{Pre-supernova feedback time-scales}

We now consider three feedback mechanisms that operate during the early stages of star formation: photoionisation (i.e.~the thermal pressure of the warm gas), stellar winds (mass loss from the stars), and radiation pressure (from the photon field emitted by the young stars). In the following, we summarise the equations used here to determine the characteristic time-scales for GMC dispersal by each of these three early mechanisms. The full derivation can be found in \citet[and references therein]{Kruijssen2019}.

The characteristic time-scales for GMC dispersal by photoionisation, stellar winds, and radiation pressure depend on stellar region masses as well as on GMC properties, such as cloud radii, surface densities and volume densities. The cloud masses and radii (and the resulting cloud surface densities) are obtained by applying an updated version of the cloud segmentation algorithm CPROPS \citep{Rosolowsky2006} to the observed CO emission maps at their native resolution (see \citealt{Rosolowsky2020} and Hughes et al.\ in prep.\ for a detailed description of the method and its application to the full PHANGS sample). One CO-to-\HH\ conversion factor per galaxy is adopted to calculate cloud masses, depending on the metallicity following \citet{Sun2020}. We note that some of the structures identified as `clumps' by CPROPS can be as large as several hundreds of parsecs in diameter. While these may represent real giant molecular complexes or result from the finite resolution of the ALMA observations, the assumption of spherical GMCs is difficult to justify in these cases, in view of the typical gas disc scale height of star-forming galaxies ($\lesssim 100$\,pc for the Milky Way, \citealt{Dame2001, Heyer2015, Marasco2017}; $\lesssim 300$\,pc in nearby galaxies, \citealt{Yim2014, Levy2019, Wilson2019}). We therefore follow a similar approach as \citet{Rosolowsky2020} and limit the maximum vertical diameter of the clouds to the molecular gas disc scale height, which we assume here to be equal to $\lambda$ \citep{Kruijssen2019}. We then calculate the cloud volume density ($\rho_{\rm GMC}$) by assuming an ellipsoidal geometry.\footnote{We assume here a homogeneous cloud volume density. Assuming a different density distribution within clouds would only marginally affect our results.}

The characteristic time-scale for molecular cloud dispersal by photoionisation ($t_{\rm phot}$) is obtained by equating the radius evolution of the ionisation shock around a young stellar region \citep{Spitzer1978, Hosokawa2006} to the median GMC radius, $r_{\rm GMC}$, in the vertical dimension (i.e. limited by the scale height of the molecular disc): 
\begin{equation}
\label{eq:tphot}
t_{\rm phot}=\frac{4}{7}\left(\frac{3}{4}\right)^{1/2}\frac{r_{\rm S}}{c_{\rm s}}\left[\left(\frac{r_{\rm stdev}}{r_{\rm S}}\right)^{7/4}-1\right],
\end{equation}
where $c_{\rm s}$ is the sound speed in the ionised gas and we adopt $r_{\rm stdev}=r_{\rm GMC}/1.91$ to match the definition of cloud radius in our analysis \citep{Kruijssen2018}, determined as the standard deviation of a Gaussian emission peak. The Str\"{o}mgren radius, $r_{\rm S}$, is  defined as follows: 
\begin{equation}
\label{eq:rs}
r_{\rm S}=\left(\frac{3m_{\rm H}^2}{4(1.1)\pi \alpha_B X_{\rm H}^2}\frac{\dot{N}_{\rm LyC}}{\rho_{\rm GMC}^2}\right)^{1/3},
\end{equation}
where $m_{\rm H}=1.7\times10^{-24}~{\rm g}$ is the mass of the hydrogen atom, the factor 1.1 accounts for the additional free electrons present assuming that He is singly ionised \citep{Krumholz2017}, $\alpha_{\rm B}=2.56\times10^{-13}~{\rm cm}^3~{\rm s}^{-1}\times(T/10^4~{\rm K})^{-0.83}$ is the case-B recombination rate \citep{Tielens2005}, with $T$ the electron temperature, and $X_{\rm H}=0.7$ is the hydrogen mass fraction. The dependence of $t_{\rm phot}$ on the electron temperature is weak, so we assume a constant, typical value of $T = 10^4$\,K \citep[e.g.][]{Kennicutt2003b}. 
The sound speed is defined as:
\begin{equation}
c_{\rm s}  = \left( \frac{k_{\rm B}T}{\mu m_{\rm H}} \right)^{1/2},
\end{equation}
where $k_{\rm B} = 1.381\times 10^{-23}$\,m$^2$\,kg\,s$^{-2}$\,K$^{-1}$ is the Boltzmann constant and $\mu = 0.62$ is the mean molecular weight in ionised gas.
The Lyman continuum photon emission rate is defined as $\dot{N}_{\rm LyC}/{\rm s}^{-1}=10^{46.5}(M_\text{H{\sc ii}}/\msun)$ for an \HII\ region of stellar mass $M_\text{H{\sc ii}}$ \citep{Leitherer14}.\footnote{Strictly speaking, this value is valid for the zero-age main sequence. However the change in the Str\"{o}mgren radius after 4\,Myr due to the evolution of the Lyman continuum photon emission rate is less than 50 per cent for non-rotating stellar models, and much smaller for rotating models. We therefore assume this value to be constant over the duration of the feedback phase.} In principle, equation~\ref{eq:tphot} applies to the early stages of the expansion. We assume here for simplicity that this solution is valid for the entire duration $t_{\rm phot}$, although we note that more complex analytical solutions exist \citep[e.g.][]{Williams2018}. In addition, the calculated value can be a lower limit to the true time-scale in cases where the effects of self-gravity cannot be neglected.

\begin{table*}
\begin{center}
\begin{tabular}{lllllllll}
\hline
Galaxy & \tHa & \tover & \tsn & $t_{\rm phot}$ & $t_{\rm wind}$ & $t_{\rm cool}$ & $t_{\rm rad}$ & $t_{\rm comb}$ \\
\hline
NGC0628 & $7.6^{+0.7}_{-0.5}$ & $3.2^{+0.6}_{-0.4}$ & $8.4$ & $1.6$ & $\mathbf{1.4}$ & $2.2$ & $5.9$ & $0.7$ \\
NGC3351 & $6.8^{+0.8}_{-0.7}$ & $2.5^{+0.8}_{-0.6}$ & $8.9$ & $\mathbf{1.8}$ & $1.9$ & $1.6$ & $6.9$ & $0.8$ \\
NGC3627 & $7.2^{+0.8}_{-0.8}$ & $2.8^{+0.8}_{-0.7}$ & $3.9$ & $\mathbf{1.2}$ & $1.2$ & $4.3$ & $3.8$ & $0.5$ \\
NGC4254 & $9.2^{+1.2}_{-1.0}$ & $4.8^{+1.1}_{-1.0}$ & $4.1$ & $\mathbf{1.8}$ & $1.9$ & $4.7$ & $6.3$ & $0.8$ \\
NGC4303 & $8.4^{+1.8}_{-1.1}$ & $4.0^{+1.7}_{-1.0}$ & $4.2$ & $2.4$ & $\mathbf{1.9}$ & $4.3$ & $6.9$ & $0.9$ \\
NGC4321 & $7.6^{+0.8}_{-0.7}$ & $3.3^{+0.7}_{-0.6}$ & $4.2$ & $\mathbf{1.3}$ & $1.6$ & $3.9$ & $5.3$ & $0.6$ \\
NGC4535 & $8.3^{+1.2}_{-0.9}$ & $3.9^{+1.2}_{-0.9}$ & $4.3$ & $\mathbf{1.1}$ & $1.3$ & $4.4$ & $4.7$ & $0.5$ \\
NGC5068 & $5.5^{+0.5}_{-0.3}$ & $1.0^{+0.4}_{-0.3}$ & $9.8$ & $1.2$ & $\mathbf{0.7}$ & $1.6$ & $3.6$ & $0.4$ \\
NGC5194 & $9.0^{+2.1}_{-1.2}$ & $4.8^{+2.1}_{-1.1}$ & $4.2$ & $\mathbf{3.2}$ & $8.7$ & $0.5$ & $5.6$ & $1.6$\\
\hline
\end{tabular}
\caption{Summary of the average observed and predicted feedback time-scales (in Myr) for each galaxy of the sample. The 1$\sigma$ uncertainties are indicated for the observed duration of the stellar phase, \tHa , and of the feedback phase. The minimum predicted feedback time-scale between supernova delay time $t_{\rm SN}$, photoionisation $t_{\rm phot}$, winds $t_{\rm wind}$ and radiation pressure $t_{\rm rad}$ is indicated in boldface. The cooling time $t_{\rm cool}$ and the combined predicted time-scale $t_{\rm comb}$ are also presented (see text).}
\label{tab:timescales}
\end{center}
\end{table*}
The characteristic time-scale for molecular cloud dispersal by stellar winds ($t_{\rm wind}$) is obtained by equating the radius evolution of the energy-driven wind shock \citep{Weaver1977} to the median GMC radius in the vertical dimension:
\begin{equation}
\label{eq:twind}
    t_{\rm wind} = \begin{cases}
    \left(\frac{154\pi}{125}\frac{\rho_{\rm GMC}}{L_{\rm wind}}\right)^{1/3} r_{\rm stdev}^{5/3}, & \text{if  $t_{\rm wind} \leq t_{\rm cool}$}\\
    \left(\frac{154\pi}{125}\frac{\rho_{\rm GMC}}{L_{\rm wind}}\right)^{4/5} r_{\rm stdev}^{4} t_{\rm cool}^{-7/5}, & \text{otherwise}.
    \end{cases}
\end{equation}
where $L_{\rm wind}/{\rm erg}~{\rm s}^{-1}=10^{34}\times(M_\text{H{\sc ii}}/\msun)$ is the mechanical luminosity of the wind driven by a stellar mass $M_\text{H{\sc ii}}$ \citep{Leitherer14}, and $t_{\rm cool}$ is the cooling time of the hot bubble \citep{MacLow1988}, given by:
\begin{equation}
    \frac{t_{\rm cool}}{\myr}=0.96\left(\frac{Z}{{\rm Z}_\odot}\right)^{-35/22}\left(\frac{L_{\rm wind}}{10^{37}~{\rm erg}~{\rm s}^{-1}}\right)^{3/11}\left(\frac{\rho_{\rm GMC}}{20~\cmc}\right)^{-8/11}.
\end{equation}
Variations of the wind luminosity during the feedback phase are small \citep{Leitherer14} and we assume a constant value throughout this phase. We note that the above calculation neglects energy leakage by shell fragmentation \citep[e.g.][]{Rahner2019} and is accurate to within a factor of two relative to results obtained with detailed cooling functions \citep[e.g.][]{MacLow1988}. We also neglect the momentum input from the wind, which would become important in the case where $t_{\rm wind} > t_{\rm cool}$ \citep[see e.g.][]{Silich2013, Rahner2017, Rahner2019}, and would lead to overestimating $t_{\rm wind}$. This might have a significant impact only for NGC5194, and even then it is unlikely to change our conclusions, given the large difference between $t_{\rm phot}$ and $t_{\rm wind}$ (3.2\,Myr and 8.7\,Myr, respectively, see Table~\ref{tab:timescales}).

Finally, following \citet{Kruijssen2019}, the characteristic time-scale for molecular cloud dispersal by radiation pressure ($t_{\rm rad}$) is obtained by equating the radius evolution of the radiation pressure-driven shell to the median GMC radius in the vertical dimension:
\begin{equation}
\label{eq:trad}
t_{\rm rad}=\left(\frac{2\pi c}{3}\frac{\rho_{\rm GMC}}{L_{\rm bol}}\right)^{1/2} r_{\rm stdev}^2, 
\end{equation}
where $L_{\rm bol}/{\rm erg}~{\rm s}^{-1}=10^{36.6}\times(M_\text{H{\sc ii}}/\msun)$ is the bolometric luminosity of a solar metallicity stellar population \citep{Leitherer14} and $c$ is the speed of light. This assumes that GMCs are optically thin to infrared radiation, following the conclusions from \citet{Kruijssen2019} for GMCs similar to the ones considered in our sample. 

The resulting analytically-predicted time-scales for these three feedback mechanisms are shown in Figure~\ref{fig:profiles} for each radial bin. In order to determine the above time-scales, we implicitly assume that there is perfect coupling between winds and photons from the young \HII\ regions and the ambient ISM. This is not necessarily the case and in fact the efficiency with which the feedback mechanisms couple with the surrounding ISM is likely smaller than unity. This is discussed in Section~\ref{sec:efficiency}. 

      \subsubsection{Combining feedback mechanisms}

\begin{figure*}
	\includegraphics[width=\linewidth]{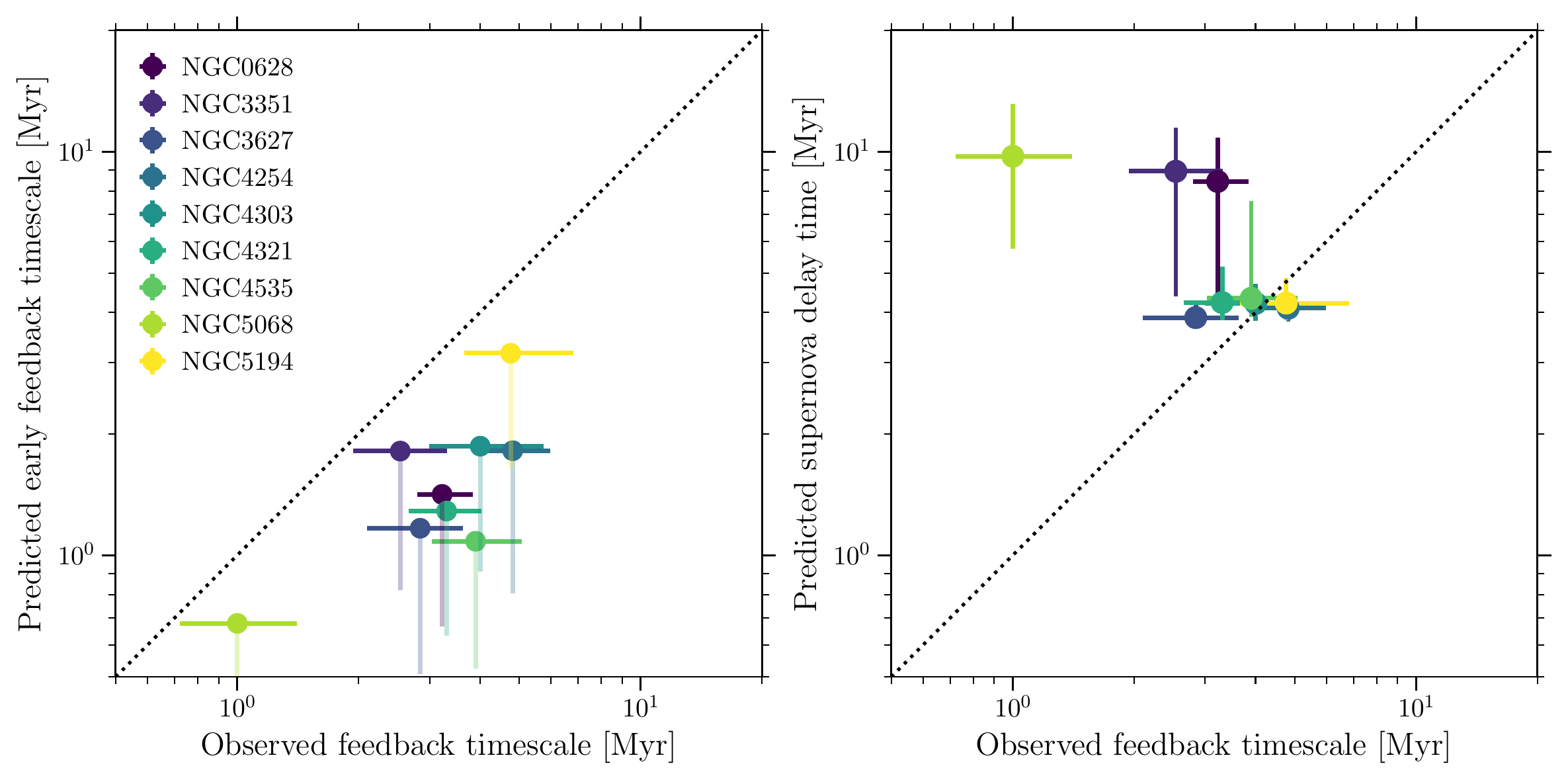}
    \caption{\textit{Left panel:} Predicted early feedback time-scale (ignoring the effects of supernovae) versus observed feedback time-scale \tover\ for the full galaxies in our sample. The solid circles indicate the predicted time-scale considering only the dominant feedback mechanism (with the shortest time-scale, $t_{\rm min}$), with the horizontal error bars representing the 1$\sigma$ uncertainties on \tover . The shaded vertical error bars indicate the range between \tcomb\ and \tmin . All data points lie below the 1:1 relation (dashed line), implying that the coupling between stellar feedback and the surrounding molecular gas is not perfect. 
    \textit{Right panel:} Supernova delay time versus observed feedback time-scale. In galaxies above the 1:1 dotted line, the feedback time-scale is smaller than the supernova delay time. In these galaxies, GMCs are destroyed predominantly by photoionisation and winds. In galaxies below the line, the feedback time-scale is longer than (but still consistent with) the supernova delay time. The comparison between both panels provide an unambiguous interpretation of our measurements, which is that GMCs are generally dispersed by early, pre-supernova feedback.
    }
    \label{fig:tfb_match}
\end{figure*}

In the extreme case, if we assume that all of the above pre-supernova feedback mechanisms act simultaneously and combine in a constructive way, we can define the minimum combined time-scale as the inverse of the sum of all dispersal rates:
\begin{equation}
    t_{\rm comb} = (t_{\rm phot}^{-1} + t_{\rm wind}^{-1} + t_{\rm rad}^{-1})^{-1}.
    \label{eq:tcomb}
\end{equation}
This expression does not include the supernova delay time, because we consider supernova explosions as an instantaneous mechanism that destroys the cloud, rather than a continuous process affecting the cloud over a finite characteristic time-scale. In practice, the minimum of \tcomb\ and \tsn\ sets a lower limit on the expected feedback time-scale, under the assumption that all mechanisms interact constructively and simultaneously to disperse the parent molecular cloud. It also provides a lower limit if not all early feedback mechanisms act simultaneously and different mechanisms dominate at different points during the expansion of the shell. This lower limit is shown as the lower boundary of the blue-shaded area in Figure~\ref{fig:profiles}. 

It is not clear if the above three feedback mechanisms combine constructively in nature \citep[see e.g.][]{Rahner2017}. The opposite extreme is that the mechanisms are fully independent and that only the mechanism with the shortest time-scale is responsible for cloud dispersal. In that case, the feedback time-scale is simply set by the minimum dispersal time-scale across the three mechanisms,
\begin{equation}
    t_{\rm min} = \min(t_{\rm phot},t_{\rm wind},t_{\rm rad}) .
    \label{eq:tmin}
\end{equation}
In practice, the appropriate time-scale to compare with the observed feedback time-scale is likely to be located somewhere in between these two extremes. 

As pointed out in Section~\ref{sec:environment}, the limited size of our sample after excluding radial bins potentially affected by biases do not allow us to identify variations of the feedback time-scale within galaxies. In the following, we will therefore only consider the galaxy-wide measurements, obtained across the entire field of view. Table~\ref{tab:timescales} summarises all observed and predicted time-scales for each galaxy in the sample.

In Figure~\ref{fig:tfb_match}, we compare the measured galaxy-wide feedback time-scales to the predicted early feedback time-scales (equations~\ref{eq:tcomb} and~\ref{eq:tmin}, ignoring the effect of supernovae) and the supernova delay time. We find that $t_{\rm min}$  (solid data points) and \textit{a fortiori} \tcomb\ (lower end of the transparent vertical error bar) are lower than the observed \tover . This implies that the early feedback mechanisms considered here are capable of disrupting the parent GMC. It also indicates that the stellar feedback mechanisms are likely not maximally efficient in dispersing the molecular gas. We estimate the implied coupling efficiency of the dominant feedback mechanism with the surrounding interstellar medium (ISM) for each galaxy in Section~\ref{sec:efficiency}.  

In the right-hand panel of Figure~\ref{fig:tfb_match}, we compare the measured feedback time-scale to the supernova delay time for the full galaxies. We note that the predicted supernova delay time mostly exceeds the feedback time-scale. For NGC4254 and NGC5194, the supernova delay time is smaller than the observed feedback time-scale, such that supernova explosions could contribute to cloud dispersal. None the less, the time-scales for early feedback are shorter than the supernova delay time for all galaxies in the sample. At face value, the comparison of both panels in Figure~\ref{fig:tfb_match} thus indicates that early, pre-supernova feedback dominates GMC dispersal in the majority of environments considered here. This conclusion could be nuanced if there exists an embedded phase of massive star formation, during which \Ha\ is undetectable. We discuss this possibility in Section~\ref{sec:interpretation}.

\section{Discussion}
\label{sec:discussion}

\subsection{Feedback coupling efficiency}
\label{sec:efficiency}

\begin{figure*}
	\includegraphics[width=\linewidth]{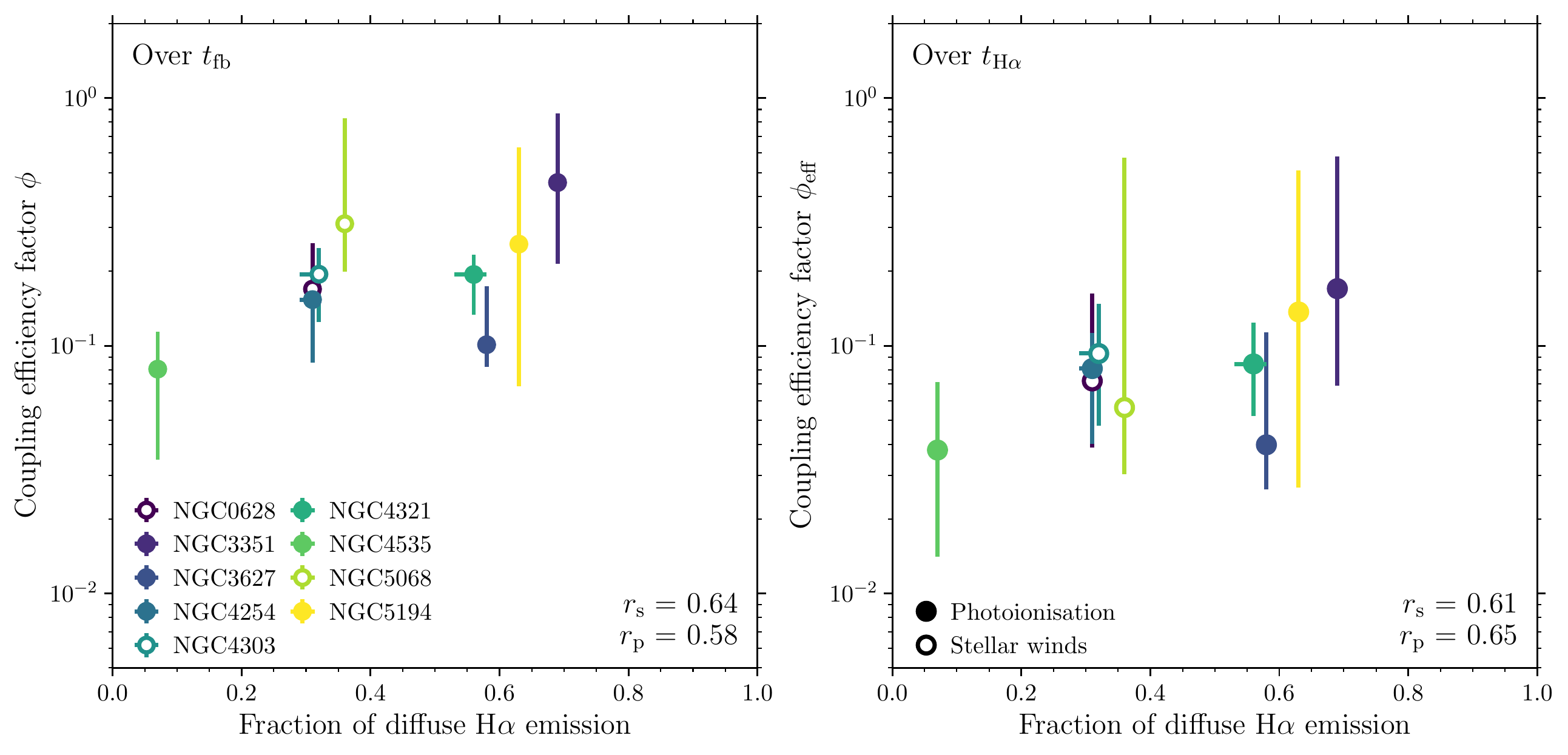}
    \caption{Feedback coupling efficiency integrated over the duration of the feedback phase ($\phi$, left panel) and over the young stellar region lifetime ($\phi_{\rm eff}$, right panel), as a function of the fraction of diffuse \Ha\ emission (see the text). The filled circles indicate galaxies in which photoionisation is predicted to be the dominant process for GMC destruction, whereas the open circles indicate galaxies in which stellar winds are the dominant process. We observe a tentative correlation between the (effective) coupling efficiency and the fraction of diffuse \Ha\ emission, with Spearman and Pearson correlation coefficients ($r_{\rm s}$ and $r_{\rm p}$, respectively) as indicated in each panel.
    }
    \label{fig:efficiency_obs}
\end{figure*}

Equations~\ref{eq:tphot}-\ref{eq:trad} assume that there is complete coupling between winds and photons in the young \HII\ regions and the ambient ISM in order to estimate the time-scales for cloud dispersal by photoionisation, stellar winds and radiation pressure. In practice, the fact that the ISM is highly structured means that low density channels will be carved out in GMCs, through which photons and winds can escape \citep[e.g.][]{Rogers2013, Dale2015,Walch2015}. As a result, the coupling between feedback and the surrounding ISM is unlikely to be perfect, which can explain the longer observed feedback time-scales compared to the predicted ones in the case where we assume maximal coupling efficiency of the feedback (see Figure~\ref{fig:tfb_match}). Simulations show instantaneous escape fractions up to several tens of per cent, with strong temporal variations \citep[e.g.][]{Howard2017, Rahner2017, Howard2018, Kimm2019}. Based on the simulations of \cite{Kim2019}, up to $\sim$ 50 per cent of the ionising radiation could escape from $10^5$\,\msun\ clouds of low molecular gas surface density ($10-20$\,\msun\,pc$^{-2}$) within the first 3\,Myr after the onset of star formation. In their simulations, this limits the efficiency of the coupling between stellar feedback and the parent cloud and potentially results in longer characteristic time-scales for these feedback mechanisms. By contrast, they predict that only $\sim$10 per cent of the ionising radiation is expected to escape from higher mass clouds ($\sim 10^6$\,\msun ) in high gas surface density environments such as NGC5194. Here we compare the observed feedback time-scale to the minimum analytical time-scale in each galaxy, $t_{\rm min}$, to constrain the maximum coupling efficiency between feedback and the surrounding ISM.

We scale $\dot{N}_{\rm LyC}$, $L_{\rm wind}$ and $L_{\rm bol}$ in equations~\ref{eq:tphot}, \ref{eq:twind} and \ref{eq:trad} by a coupling efficiency factor $0\leq\phi\leq1$. We then solve for $\phi$ by equating \tover\ and \tmin , taking into account only the fastest dispersal mechanism. The resulting coupling efficiencies are presented in the left-hand panel of Figure~\ref{fig:efficiency_obs}, shown as a function of the fraction of diffuse \Ha\ emission measured by \cite{Chevance2020}.\footnote{This diffuse fraction is defined in Fourier space as the emission on spatial scales $\gtrsim 10 \times \lambda$, with $\lambda$ the characteristic distance between independent regions as measured with our formalism.} During the feedback phase (i.e.\ when CO and \Ha\ emission coexist), we measure relatively low average coupling efficiencies, ranging between 8~per~cent (for NGC4535) and 55~per~cent (for NGC5068), for the dominant feedback mechanism (acting on the shortest time-scale) in each galaxy. The uncertainties on the coupling efficiencies are propagated from the 1$\sigma$ uncertainties on the feedback time-scale. We reiterate that we only consider the feedback mechanism with the shortest predicted time-scale here. If several feedback mechanisms interact constructively, the predicted combined feedback time-scale in a given galaxy would be shorter (i.e.\ the offset between the observed and predicted feedback time-scales would increase) and the inferred coupling efficiency would be smaller. 

The inferred coupling efficiency depends on the time interval over which it is measured. In the left-hand panel, we only consider the feedback phase, during which (part of) the natal cloud is still present. However, if the coupling efficiency is measured over a longer time interval (e.g.\ when interested in the fate of all emitted photons, or of all photons emitted before a certain time), the effective coupling efficiency is lower, because all photons emitted after a time $t=t_{\rm fb}$ will escape. It is straightforward to describe this effect analytically. The right panel of Figure~\ref{fig:efficiency_obs} shows the effective coupling efficiency, $\phi_{\rm eff}$, when averaging over the entire lifetime of the \HII\ regions (\tHa , see Table~\ref{tab:timescales}), which is defined as:
\begin{equation}
    \phi_{\rm eff} = \phi \times \frac{\text{\tover}}{\text{\tHa}}.
\end{equation}
The effective coupling efficiency integrated over the entire \HII\ region lifetime ranges between 4 per cent and 17 per cent in our sample. By construction, $\phi_{\rm eff}$ is lower than $\phi$ because it takes into account the time during which the cloud is already dispersed, and the coupling is effectively zero. This effect is the most visible for NGC5068, where the coupling is high (55 per cent) during the (short) feedback phase, but drops to $\sim$ 10 per cent when integrating over the \HII\ region lifetime.

According to \citet{Conroy2012}, about 30~per~cent of all massive stars in the Galaxy are runaways, with velocities exceeding 30\,km\,s$^{-1}$. This would mean that the central \HII\ region stellar mass (equation~\ref{eq:MHII}) and the corresponding feedback luminosities (equation~\ref{eq:tphot}--\ref{eq:trad}) might be overestimated by the same percentage. While we did not correct for this effect when calculating these masses and luminosities, the possible existence of runaways is implicitly incorporated in the feedback coupling efficiencies calculated here. After all, their existence would increase the observed feedback time-scale relative to a case without runaways, and lead to a corresponding decrease of the inferred coupling efficiency. By contrast, runaways might increase the average escape fraction of ionising photons, with an estimated effect of up to $\sim 22$ per cent \citep{Kimm2014}. In the context of Figure~\ref{fig:efficiency_obs}, this might imply that the data points are shifted towards the bottom right by up to a maximum of $\{\Delta x,\Delta y\}=\{0.22, 0.15~{\rm dex}\}$. The maximum vertical shift due to runaways is smaller than the uncertainties.

Recent studies show that both ionising photons leaking out of \HII\ regions and low-mass evolved stars likely contribute to the existence of a reservoir of diffuse ionised gas \citep[e.g.][]{Poetrodjojo2019}. We therefore explore the potential relation between the (effective) coupling efficiency and the fraction of diffuse \Ha\ emission in Figure~\ref{fig:efficiency_obs}. We observe a tentative correlation in both panels.  If true, it is not clear where this correlation might come from. The opposite correlation would be expected if the diffuse \Ha\ emission on large scales in galaxies arises from photons leaking out of \HII\ regions, which did not contribute to the dispersal of the GMCs themselves \citep[e.g.][]{Mathis1986, Sembach2000, Wood2010}. Belfiore et al.\ (in prep.) argue that the \Ha\ surface density of the diffuse ionised gas is compatible with radiation leaking out of \HII\ regions in NGC4254 and NGC4535. To fully understand the relation between the diffuse \Ha\ emission and the feedback coupling efficiency, it would be necessary to compare the energy budget needed to produce the diffuse \Ha\ with the total energy leaking out of \HII\ regions given our inferred coupling efficiencies. Such a comparison would go well beyond the scope of this paper and therefore we defer it to future work.

\begin{figure*}
	\includegraphics[width=\linewidth]{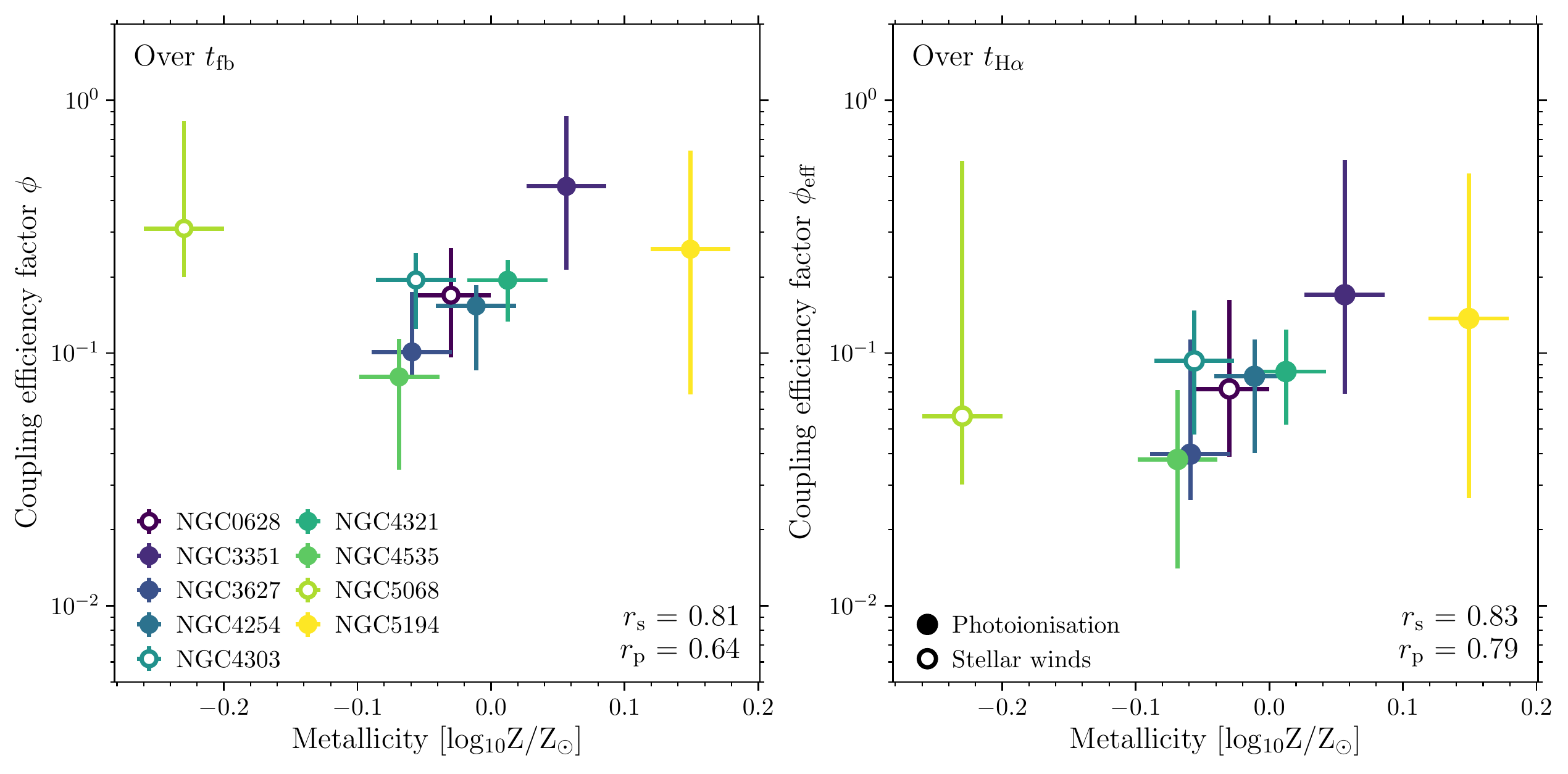}
    \caption{Same as Figure~\ref{fig:efficiency_obs}, showing the coupling efficiency integrated over the duration of the feedback phase (left) and the effective coupling efficiency integrated over the young stellar region lifetime (right) as a function of the average gas phase metallicity relative to solar. With the exception of NGC5068 (which is also excluded from the Spearman and Pearson correlation coefficients $r_{\rm s}$ and $r_{\rm p}$ indicated in each panel), the weak decrease of the (effective) coupling efficiency with decreasing metallicity would be consistent with a more porous structure of the ISM towards lower metallicities.}
    \label{fig:effective_efficiency_Z}
\end{figure*}

\begin{figure*}
	\includegraphics[width=\linewidth]{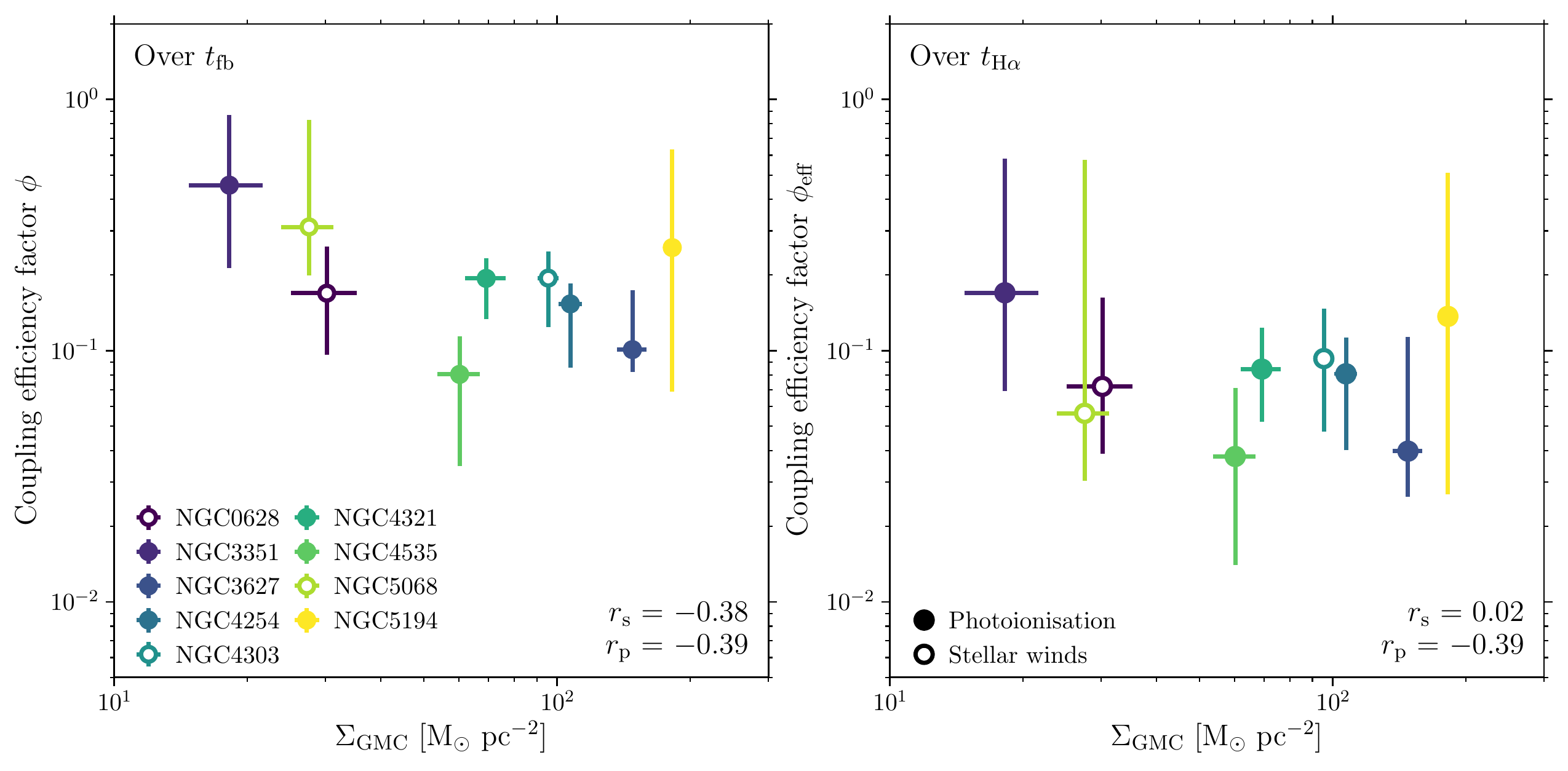}
    \caption{Same as Figure~\ref{fig:efficiency_obs}, showing the coupling efficiency integrated over the duration of the feedback phase (left) and the effective coupling efficiency integrated over the duration of the young stellar region lifetime (right) as a function of median GMC surface density in the galaxy disc measured on 120\,pc scale from \citet{Sun2018}. The Spearman and Pearson correlation coefficients ($r_{\rm s}$ and $r_{\rm p}$, respectively) are indicated in each panel. Due to the limited size of our sample and the large uncertainties on the measured coupling efficiencies, these observations do not allow us to confirm the expected trend of increasing coupling efficiency with increasing cloud surface density (see the text).}
    \label{fig:effective_efficiency_GMC}
\end{figure*}

In Figures~\ref{fig:effective_efficiency_Z} and \ref{fig:effective_efficiency_GMC}, we explore possible correlations of the (effective) coupling efficiency with the galaxy-scale averaged metallicity and the median GMC surface density ($\Sigma_{\rm GMC}$) measured at 120\,pc resolution by \citet{Sun2018}. With the exception of NGC5068, Figure~\ref{fig:effective_efficiency_Z} might indicate a weak trend of increasing coupling efficiency with increasing metallicity. This would be in agreement with the fact that the ISM becomes more porous at low metallicity, allowing for increased leakage of photons and winds through low-density channels \citep[e.g.][]{Petitpas1998, Lebouteiller2012, Cormier2015, Chevance2016}. This is also supported by simulations of low-metallicity clouds, which exhibit increased Lyman continuum escape fractions relative to solar metallicity \citep[e.g.][]{Kimm2019}. Alternatively, a feedback coupling efficiency smaller than unity might result from radiative losses \citep{Tielens2005}. It may be possible to distinguish between photon leakage and radiative cooling by considering the scaling of the coupling efficiency with GMC (volume) density. Figure~\ref{fig:effective_efficiency_GMC} shows the (effective) coupling efficiency as a function of the GMC surface density\footnote{This is evaluated at a constant physical scale of 120~pc and is used here as a proxy for the volume density following \citealt{Sun2018}.} and does not reveal any statistical correlation between both quantities. If radiative cooling were responsible for the inefficient coupling of the feedback to the surrounding ISM over the time interval \tover, we would expect the coupling efficiency to drop steeply with increasing GMC density. No such strong relation is observed, which lends support to the interpretation that photon leakage likely dominates over radiative cooling during the feedback phase. Further work, extending our analysis to regions with higher cloud surface densities, would be necessary to provide a definitive assessment of which mechanism limits the coupling efficiency.

A direct comparison with the feedback coupling efficiency (or escape fraction, $f_{\rm esc} = 1-\phi$) measured in numerical simulations is not straightforward. \cite{Kim2019} present the integrated cumulative escape fraction of ionising and non-ionising radiation over 3\,Myr after the creation of the first star particle. These authors find a strong decrease of the escape fraction (and thus an increase of the feedback coupling efficiency) with the GMC surface density, which is inconsistent with our observational measurements. However, there exist subtle differences that complicate a direct comparison. Their adopted time interval of 3\,Myr potentially includes a deeply embedded phase, during which star formation is on-going but no \Ha\ emission is visible. Including a deeply embedded phase of star formation, with a lower escape fraction and an elevated coupling efficiency, would change the average coupling efficiency that we measure in each galaxy. Likewise, taking into account the interplay between different feedback mechanisms (instead of only using the dominant one) would decrease the measured coupling efficiency. For these reasons, and because the duration of the deeply embedded phase of star formation may vary between galaxies \citep{Kim2020} and is unconstrained for the galaxies in our sample, we do not attempt a quantitative comparison with the results from \cite{Kim2019}. None the less, we note that the range of escape fractions we measure after a time \tover\ is $\sim 45-92$~per~cent, which is somewhat larger but qualitatively similar to the range of $\sim 30-60$~per~cent found by \cite{Kim2019} for $10^5$\,\msun\ clouds of similar surface densities (10-200 \msun \,pc$^{-2}$) as in the galaxies of our sample \citep[see][]{Sun2018}. Similarly, we note a relatively good agreement with the escape fraction of 63~per~cent integrated over 5\,Myr measured by \cite{Howard2018}, for a simulated cloud of $10^5$\,\msun\ with a density of 100\,cm$^{-3}$, as well as with the predictions of the WARPFIELD model \citep{Rahner2019, Pellegrini2020}. In the latter model, the cumulative escape fraction of ionising photons for a cloud resembling our observations (mass of $10^5$\,\msun\ and gas mass surface density of 100\,\msun \,pc$^{-2}$) ranges between 40 and 50~per~cent when integrated over 3\,Myr (E.~Pellegrini, private communication).

\subsection{Physical interpretation}
\label{sec:interpretation}
We have measured the duration of the evolutionary phase during which CO and \Ha\ emission coexist spatially in young star-forming regions in Section~\ref{sec:measurement}. This duration is universally short and ranges between $\sim$ 1 and 5\,Myr for our sample of star-forming disc galaxies. Although the number of radial bins in which this time-scale can be reliably determined is relatively limited in some galaxies, we find no significant trend with galactocentric radius across the fields of view probed here. The identified variations within galaxies are mostly linked to morphological features such as the presence of a bar or the co-rotation radius (and are therefore affected by methodological biases). We now discuss the physical interpretation of this co-existence time-scale.

We first emphasise that our analysis enables us to measure the lifetime of each phase as the `visibility' duration of a specific tracer. What we refer to as `GMC dispersal' therefore corresponds to the CO emission being below our detection limit (corresponding to clouds of $\sim10^5$\msun ). As a result, we cannot distinguish between a scenario where the molecular gas fragments and is displaced away from the young stellar region, remaining mostly molecular but in a more diffuse phase, and a scenario where the molecules are actually destroyed and a majority of the gas mass is removed from the molecular phase. Nevertheless, the durations of the different phases of the molecular cloud lifecycle measured here allow us to infer which physical mechanisms are at least partially responsible for driving this cycle.

The short ($10-30$\,Myr) cloud lifetime measured by \citet{Chevance2020} roughly corresponds to a GMC dynamical time-scale, which suggests that molecular cloud dispersal does not happen without associated high-mass star formation. If molecular clouds went through multiple cycles of dynamical dispersal and re-formation before forming high-mass stars, the integrated cloud lifetime (encompassing all the time spent in a CO-bright phase) would necessarily be longer than a dynamical time. We therefore conclude that the dispersal of GMCs in the observed galaxies is systematically associated with the presence of high-mass star formation.

Similarly, it is unlikely that the observed spatial decorrelation between clouds and young stellar regions results from a dynamical displacement of the entire cloud relative to the young stellar region. Ejecting young clusters from their parent clouds on such short time-scales would be inconsistent with typically measured cloud-scale velocity dispersions \citep{Sun2018}. As a result, we can directly relate the duration of the feedback phase measured here to the time-scale over which molecular clouds are dispersed (or fragmented) by feedback from high-mass stars. 

In addition to helping constrain which stellar feedback mechanisms play a major role in limiting the cloud lifetime and star formation efficiency, the short feedback time-scale measured here also helps constrain the possibility of multiple generations of high-mass star formation in GMCs. Our short feedback time-scales require multiple events of high-mass star formation to take place relatively concurrently (within \tover ) within a single GMC, but significant age spreads ($>10$\,Myr) should exist across GMC complexes \citep{Efremov1998}. Finally, the rapid destruction of GMCs in star-forming galaxies measured here constitutes a fundamental test of the feedback recipes used in galaxy simulations. \cite{Fujimoto2019} have shown that simulations may reproduce the observed instantaneous properties of cloud populations and their host galaxies, but that this does not necessarily guarantee that the observed cloud lifecycles with short feedback time-scales are reproduced as well.

A short GMC dispersal time-scale after the onset of high-mass star formation sets strong constraints on how efficiently gas is converted into stars on the cloud scale. While some studies find that the embedded phase of high-mass star formation (without associated visible \Ha\ emission) is short (i.e.\ $\lesssim 1~\myr$; see e.g.\ \citealt{Prescott2007, Hollyhead2015}, recent measurements performed by \cite{Kim2020} in a sample of five galaxies (mostly in the Local Group) reveal a typical duration of the embedded phase of star formation of $\sim 3$\,Myr. These measurements are based on the spatial distributions of CO and 24\mic\ emission peaks in galaxies, using the same analysis method as adopted here. In cases where a significant fraction of the high-mass star formation is embedded and therefore invisible in \Ha , the total duration of the feedback time-scale measured here using \Ha\ would be underestimated by the same amount. The total duration of the feedback phase could therefore become consistent with the supernova delay time for 2/3 of the galaxies in our sample (except for NGC0628, NGC3351 and NGC5068). However, there are two important factors that plausibly limit such an underestimation of the feedback time-scale. Firstly, contrary to the \Ha\ emission, which is a clear sign of ongoing or recent high-mass star formation, the 24\mic\ emission is dominated by OB-type stars, which have lifetimes up to $\sim40$\,Myr. Secondly, it is plausible that the most massive stars (with the shortest lifetimes of $\sim4$\,Myr) finish their formation process only quite late during the embedded phase \citep[e.g.][]{Tan2014,Cyganowski2017}, upon which they rapidly cause the region to become unembedded \citep[also see][]{Barnes2020}. Their explosions as supernovae then follow after the usual delay time, as assumed in this paper. In conclusion, while it remains possible that some high-mass stars could form at the beginning of the embedded phase and therefore have the time to explode as supernova within their parent CO cloud, it remains likely that in most cases early feedback mechanisms (winds and photoionisation) have largely dispersed the molecular gas before that time.

\section{Conclusions}
\label{sec:conclusion}

We present a systematic measurement of the duration of the feedback phase, between the onset of high-mass star formation visible in \Ha\ and the dispersal of the CO-emitting molecular cloud, across a sample of nine nearby star-forming disc galaxies. We use the statistical method presented by \cite{KL14} and \cite{Kruijssen2018}, as applied in \cite{Chevance2020}.

Across our sample, we measure relatively short durations of the feedback phase, with \tover\ ranging between $\sim$ 1 and 5\,Myr on average across the nine galaxies. Within these galaxies, variations of the feedback time-scale can mostly be attributed to morphological features such as the presence of a bar or the co-rotation radius, which are likely to bias our measurements. With these exceptions, we do not detect a significant variation of the feedback time-scale as a function of galactocentric radius within galaxies.

The short duration of the measured feedback phase and the comparison with analytical predictions strongly suggest that early feedback mechanisms (mainly photoionisation and stellar winds) efficiently disperse molecular clouds, prior to supernova explosions. By contrast, our results indicate that radiation pressure does not play a major role in GMC dispersal, in agreement with the conclusions of previous observational studies in the LMC, SMC and the nearby star-forming galaxy NGC300 \citep{Lopez2014, Kruijssen2019, McLeod2019, McLeod2020}, as well as theoretical predictions \citep[e.g.][]{KrumholzThompson2012, KrumholzThompson2013, Reissl2018}. While it is possible that the durations of the feedback phase measured here are underestimated due to the presence of an embedded phase of star formation \citep{Kim2020}, we would still expect this conclusion to hold in most cases because high-mass stars likely form late during the star formation phase. We report a correlation of the feedback time-scale with the galaxy stellar mass (as well as weaker correlations with the gas phase metallicity and with the median GMC surface density, which are likely driven by the correlation with the stellar mass), and provide a power law fit (equation~\ref{eq:fit}).

The comparison between the observed feedback time-scale and the predicted dominant feedback time-scale allows us to evaluate the efficiency of the coupling between this feedback mechanism (photoionisation or stellar winds, excluding supernovae) and the surrounding ISM. Across our galaxy sample, we find average coupling efficiencies of $\sim$ 8-55 per cent during the feedback phase, or $\sim 4-17$ per cent when integrated over the entire lifetime of the \HII\ regions (the `effective' coupling efficiency). We note that these values are likely to decrease if several mechanisms combine constructively to disperse the GMCs. We find a tentative correlation between the (effective) coupling efficiency and the gas phase metallicity, but no correlation with the median GMC surface density, contrary to the simulations by \cite{Kim2019}. However, we neglect a possible, deeply embedded phase of star formation, during which \Ha\ is completely obscured \citep{Kim2020}. The uncertainties surrounding the early feedback process during the embedded phase make it challenging to quantitatively compare our observations with simulations. 

While the limited size of our sample prohibits making a definitive assessment of how the feedback time-scale and related quantities depend on large-scale galactic properties, it is clear from the observations presented here that pre-supernova feedback mechanisms (photoionisation and stellar winds) dominate GMC dispersal in most environments. An accurate description of the GMC-scale baryon cycle in simulations of galaxy formation and evolution thus {\it requires} including the effects of these mechanisms. In our upcoming analysis of the complete PHANGS-ALMA survey, we plan to extend our reported, tentative correlations between GMC-scale feedback and the galactic environment to a statistically representative sample.

\section*{Acknowledgements}
The authors thank Eve Ostriker for helpful discussions and Eric Pellegrini for computing the integrated ionising photon escape fractions predicted by WARPFIELD.
MC and JMDK gratefully acknowledge funding from the Deutsche Forschungsgemeinschaft (DFG, German Research Foundation) through an Emmy Noether Research Group (grant number KR4801/1-1). MC, JMDK and JJK gratefully acknowledge funding from the DFG Sachbeihilfe (grant number KR4801/2-1). JMDK gratefully acknowledges funding from the European Research Council (ERC) under the European Union's Horizon 2020 research and innovation programme via the ERC Starting Grant MUSTANG (grant agreement number 714907). MC, JMDK and MRK acknowledge support from the Australia-Germany Joint Research Cooperation Scheme (UA-DAAD, grant number 57387355). MRK acknowledges support from the Australian Research Council through its Future Fellowship (FT180100375) and Discovery Projects (DP190101258) funding schemes.
AH was supported by the Programme National Cosmology et Galaxies (PNCG) of CNRS/INSU with INP and IN2P3, co-funded by CEA and CNES. 
AH and JP acknowledge support from the Programme National `Physique et Chimie du Milieu Interstellaire' (PCMI) of CNRS/INSU with INC/INP co-funded by CEA and CNES.
SCOG and RSK acknowledge support from the DFG via the Collaborative Research Center (SFB 881, Project-ID 138713538) `The Milky Way System' (sub-projects A1, B1, B2 and B8) and from the Heidelberg cluster of excellence `STRUCTURES: A unifying approach to emergent phenomena in the physical world, mathematics, and complex data', funded by the German Excellence Strategy (grant EXC-2181/1 - 390900948).
The work of AKL and JS is partially supported by the National Science Foundation under Grants No. 1615105, 1615109, and 1653300.
ER acknowledges the support of the Natural Sciences and Engineering Research Council of Canada (NSERC), funding reference number RGPIN-2017-03987.
ES, CMF, DL and TS acknowledge funding from the ERC under the European Union's Horizon 2020 research and innovation programme (grant agreement No. 694343).
ATB and FB acknowledge funding from the European Union's Horizon 2020 research and innovation programme (grant agreement No 726384/EMPIRE).
RSK thanks for funding from the European Research Council via the ERC Synergy Grant ECOGAL (grant 855130).
CMF is supported by the National Science Foundation under Award No. 1903946. 
K.K.\ gratefully acknowledges funding from the DFG in the form of an Emmy Noether Research Group (grant number KR4598/2-1, PI Kreckel).
MQ acknowledges support from the research project  PID2019-106027GA-C44 from the Spanish Ministerio de Ciencia e Innovaci\'{o}n.
The work of JS is partially supported by the National Aeronautics and Space Administration (NASA) under ADAP grants NNX16AF48G and NNX17AF39G.

This work was carried out as part of the PHANGS collaboration.
This paper makes use of the following ALMA data: 
ADS/JAO.ALMA\#2012.1.00650.S, 
ADS/JAO.ALMA\#2015.1.00925.S, 
ADS/JAO.ALMA\#2015.1.00956.S, 
ALMA is a partnership of ESO (representing its member states), NSF (USA) and NINS (Japan), together with NRC (Canada), MOST and ASIAA (Taiwan), and KASI (Republic of Korea), in cooperation with the Republic of Chile. The Joint ALMA Observatory is operated by ESO, AUI/NRAO and NAOJ.
This paper includes data gathered with the 2.5 meter du Pont Telescope located at Las Campanas Observatory, Chile, and data based on observations carried out at the MPG 2.2m telescope on La Silla, Chile.

\section*{Data availability}
The data underlying this article will be shared on reasonable request to the corresponding author.




\bibliographystyle{mnras}
\bibliography{feedback} 







\vspace{4mm}

\noindent {\it
$^{1}$Astronomisches Rechen-Institut, Zentrum f\"ur Astronomie der Universit\"at Heidelberg, M\"onchhofstrasse 12-14, D-69120 Heidelberg, Germany\\
$^{2}$Research School of Astronomy and Astrophysics - The Australian National University, Canberra, ACT, 2611, Australia\\
$^3$Centre of Excellence for Astronomy in Three Dimensions (ASTRO-3D), Canberra, ACT 2611, Australia\\
$^{4}$Institut fur Theoretische Astrophysik, Zentrum f\"ur Astronomie, Universit\"at Heidelberg, D-69120 Heidelberg, Germany\\
$^{5}$Max Planck Institute for Astronomy, Konigstuhl 17, D-69117 Heidelberg, Germany\\
$^{6}$International Centre for Radio Astronomy Research, University of Western Australia, 7 Fairway, Crawley, 6009, WA, Australia\\
$^{7}$CNRS, IRAP, 9 Av. du Colonel Roche, BP 44346, F-31028 Toulouse cedex 4, France\\
$^{8}$Universit\'{e} de Toulouse, UPS-OMP, IRAP, F-31028 Toulouse cedex 4, France\\
$^{9}$Universit\"{a}t Heidelberg, Zentrum f\"{u}r Astronomie, Institut f\"{u}r Theoretische Astrophysik, Albert-Ueberle-Str. 2, D-69120 Heidelberg, Germany\\
$^{10}$IRAM, 300 rue de la Piscine, F-38406 Saint Martin d'H\`eres, France\\
$^{11}$ Department of Astronomy, The Ohio State University, 140 West 18th Ave, Columbus, OH 43210, USA\\
$^{12}$Sorbonne Universit\'e, Observatoire de Paris, Universit\'e PSL, CNRS, LERMA, F-75005, Paris, France\\
$^{13}$Departamento de Astronom\'{i}a, Universidad de Chile, Casilla 36-D, Santiago, Chile\\
$^{14}$4-183 CCIS, University of Alberta, Edmonton, Alberta, Canada\\
$^{15}$Max-Planck Institut f\"ur Extraterrestrische Physik, Giessenbachstra\ss e 1, 85748 Garching, Germany\\
$^{16}$Argelander-Institut f\"ur Astronomie (AIfA), Universit\"at Bonn, Auf dem H\"ugel 71, D-53121 Bonn, Germany\\
$^{17}$The Observatories of the Carnegie Institution for Science, 813 Santa Barbara Street, Pasadena, CA 91101, USA\\
$^{18}$European Southern Observatory, Karl-Schwarzschild-Stra{\ss}e 2, D-85748 Garching bei M\"{u}nchen, Germany\\
$^{19}$Univ.\ Lyon, Univ.\ Lyon1, ENS de Lyon, CNRS, Centre de Recherche Astrophysique de Lyon UMR5574, F-69230 Saint-Genis-Laval France\\
$^{20}$Department of Astronomy, University of Massachusetts Amherst, 710 North Pleasant Street, Amherst, MA 01003, USA\\
$^{21}$Universit\"{a}t Heidelberg, Interdisziplin\"{a}res Zentrum f\"{u}r Wissenschaftliches Rechnen, Im Neuenheimer Feld 205, D-69120 Heidelberg, Germany\\
$^{22}$Astrophysics Research Institute, Liverpool John Moores University, IC2, Liverpool Science Park, 146 Brownlow Hill, Liverpool L3 5RF, UK\\
$^{23}$Sterrenkundig Observatorium, Universiteit Gent, Krijgslaan 281 S9, B-9000 Gent, Belgium\\
$^{24}$Observatorio Astron{\'o}mico Nacional (IGN), C/Alfonso XII 3, Madrid E-28014, Spain
}

\bsp	
\label{lastpage}
\end{document}